\documentclass{article}
\usepackage{graphicx}  
\usepackage{amsmath}   
\usepackage[compress]{cite}
\usepackage{amssymb}   
\usepackage{bm} 
\usepackage{dcolumn}
\usepackage{color}
\usepackage{mathrsfs}
\usepackage{amsfonts}
\usepackage{varioref}
\usepackage{textcomp}
\RequirePackage[colorlinks,citecolor=blue,urlcolor=magenta,linkcolor=blue]{hyperref}
\allowdisplaybreaks
\labelformat{section}{Section #1} 
\labelformat{subsection}{Section #1} 
\labelformat{subsubsection}{Section #1}
\labelformat{subsubsubsection}{Section #1}
\labelformat{equation}{Eq.~(#1)} 
\labelformat{figure}{Fig.~#1} 
\labelformat{subfigure}{Fig.~\thefigure#1} 
\labelformat{table}{Tab.~#1} 
\labelformat{appendix}{Appendix #1}
\title{Softly broken conformal symmetry with higher curvature terms}
\author{Sumanta Chakraborty\footnote{sumantac.physics@gmail.com}\\
{\small{School of Mathematical and Computational Sciences}}\\
{\small{and}}\\
{\small{School of Physical Sciences}}\\
{\small{Indian Association for the Cultivation of Science, Kolkata-700032, India}}}
\begin{document}
  
\maketitle
\begin{abstract}
In a scalar-coupled-gravity model, the quadratically divergent counter term appearing in the mass renormalization of the scalar fields must inherit corrections arising out of gravitational interactions. In this work we have explicitly demonstrated that there are no such corrections of gravitational origin to the quadratic divergences in the mass counter terms. This statement holds true irrespective of the nature of the gravitational interaction, i.e., whether gravity is described by general relativity or f(R) theory. Interestingly, it also turns out that the one loop effective action of scalar-coupled-gravity system will be well-behaved \emph{if and only if} the $f(R)$ theory is free from ghosts. In particular, the results derived in the context of f(R) theory are shown to be in exact agreement with the corresponding results derived from the equivalent scalar-tensor representation. Our analysis suggests the tantalizing possibility that the masses of the scalar fields can be consistently kept smaller than some Ultra Violet (UV) cutoff scale and is independent of the nature of the gravity theory, which may involve higher curvature corrections. All these will be true provided the matter fields and the gravity theory can be embedded consistently into a UV complete theory at the Planck scale.
\end{abstract}
\section{Introduction}

The discovery of the Higg's Boson at the Large Hadron Collider (LHC) has provided the missing bit of the Standard Model, thus cementing its place as the most successful model describing the micro world \cite{Aad:2012tfa,Chatrchyan:2012ufa,Spira:1995rr,DeRujula:2010ys,Weinberg:1967tq,Salam:1968rm,Cornwall:1973tb,Barate:2003sz,Evans:2008zzb}. However, it has also opened up the Pandora's box, paving way for several intriguing questions to emerge into the limelight. In particular, the apparent discrepancy between the electroweak ($m_{\rm EW}$) and the Planck ($m_{\rm Pl}$) scale requires an immediate answer. This discrepancy, which originates from the very small ratio $(m_{\rm EW}/m_{\rm Pl})\sim 10^{-17}$, requires an abnormal fine tuning in order to arrive at the observed value for the Higg's mass. This fine tuning problem is also known as the gauge hierarchy problem and has been one of the key research direction, in the arena of theoretical high energy physics, for the last decade (see \cite{Susskind:1982mw} and the references therein). There have been several proposals, of very different kind and sometimes exotic, to resolve this issue. These include low energy supersymmetry \cite{Martin:1997ns,Freedman:2012zz}, technicolor \cite{Appelquist:1997fp,Hill:2002ap} and spatial extra dimensions \cite{Randall:1999ee,ArkaniHamed:1998rs,Antoniadis:1990ew}, among others. However, as the recent LHC data suggests, there have been no sign whatsoever, in favour of any of these models \cite{Autermann:2017chm,Alwall:2008ag,Khachatryan:2016epu,Autermann:2016les,Shmatov:2007mg,Chatrchyan:2011fq,Belyaev:2013ida}. This motivates the suggestion that any solution to the gauge hierarchy problem will possibly deviate very little from the Standard Model of particle physics. Following which an alternative method, based on the implementation of approximate conformal symmetry in the effective low energy theory, has recently been invoked in order to avoid the gauge hierarchy problem without deviating much from the Standard Model \cite{Meissner:2007xv,Chankowski:2014fva,Meissner:2018mvq}. 

The idea of approximate conformal symmetry, or softly broken conformal symmetry can be explained along the following lines. One starts from the assumption that there exists some UV finite theory, which inherits a satisfactory resolution of the gauge hierarchy problem. The UV finite theory also introduces some distinguished UV scale $\Lambda$ ($\sim$ Planck scale $m_{\rm Pl}$). Integrating out the degrees of freedom with energy scales greater than $\Lambda$ in the UV finite theory, one arrives at a low energy effective action, which presumably will resemble the Standard Model to a very good accuracy. The classical conformal symmetry is broken in the Standard Model by the presence of the mass term in the Lagrangian for the Higg's field, but only \emph{weakly}. Since the mass of the Higg's field is much small compared to the UV cutoff scale $\Lambda$. In the context of softly broken conformal symmetry one not only demands the bare mass parameters to be small compared to the distinguished UV scale $\Lambda$, but also requires the cancellation of the quadratic divergences arising from the counter terms in the renormalized mass scales of the theory. It must be emphasized that in the perturbative approach considered here, such a cancellation of the quadratic divergences has to be performed by taking into account all the loop orders. In other words, we must \emph{add} all the contributions to the quadratic divergences arising out of all the loop orders together and then shall adjust the bare couplings accordingly, so that the quadratic divergences vanish. Since it is very difficult to compute higher loop effects in an interacting theory in an explicit manner, one generically  demonstrates the cancellation of quadratically divergent term at the one loop order and then higher loop contributions are ascertained to be small \cite{Loebbert:2018xsd,Kwapisz:2017vjt,Lewandowski:2017wov,Latosinski:2015pba}. Since the two loop effects depend primarily on the quadratic of the one loop effect, it is expected that vanishing of quadratic divergence at one loop order will keep the two loop contribution to be small, which can be cancelled by slight modification of the bare parameters \cite{Chankowski:2014fva}. Furthermore, since the bare couplings do not appear in the physical processes, adjusting the same to cancel the quadratic divergences will not affect the physics of the system in any way. Hence for the theory at energy scale $\Lambda$, the fields will effectively be massless and conformal symmetry will be (weakly) respected. Note that the above argument requires the matter fields to be renormalizable and the bare couplings are related to the running couplings at the scale $\Lambda$, i.e., $\lambda_{\rm bare}=\lambda(\Lambda)$. Therefore, in the context of softly broken conformal symmetry, the physical masses can be kept as small as one desires in a perturbative treatment as their quantum corrections has no quartic or quadratic divergences depending on the cut-off scale $\Lambda$. Rather they may have a weak Logarithmic dependence on the cutoff \cite{Chankowski:2014fva,Meissner:2018mvq,Hamada:2012bp,Meissner:2009gs,Antipin:2013exa}. 

It has been demonstrated recently in \cite{Meissner:2018mvq} that the physical masses can be kept small enough in a self-consistent manner even if the gravitational perturbations originating from the Einstein-Hilbert action are taken into account. However, near the Planck scale it is not at all justified to use simply the Einstein-Hilbert term to describe the gravitational dynamics, rather one should take into account higher curvature corrections as well. There can be several possibilities for such higher curvature corrections, to be added to the Einstein-Hilbert action. Restricting the attention to those theories for which Ostrogradsky's instability can be avoided \cite{Woodard:2015zca}, it turns out that there are only a handful of such correction terms to the gravitational action. These include, $f(R)$ gravity (for reviews see, \cite{Sotiriou:2008rp,Nojiri:2010wj,DeFelice:2010aj} and for applications in the context of the gauge hierarchy problem see, \cite{Chakraborty:2020ktp,Chakraborty:2016gpg,Carames:2012gr,Haghani:2012zq,Chakraborty:2015taq,Chakraborty:2014xla}), Gauss-Bonnet term or, in general, the full Lanczos-Lovelock series (for various geometrical aspects, see \cite{Padmanabhan:2013xyr,Dadhich:2008df,Kastor:2012se} while thermodynamical aspects have been discussed in \cite{Chakraborty:2015wma}) and the Horndeski theories \cite{Charmousis:2014mia,Barausse:2015wia,Babichev:2016rlq,Bhattacharya:2016naa,Mukherjee:2017fqz}. In four spacetime dimensions, the only non-trivial dynamics due to the higher curvature terms is from the $f(R)$ theories of gravity (Horndeski theories include additional scalar fields and thus will further complicate the situation). The Lovelock Lagrangians will make contribution only in the presence of higher spacetime dimensions. As the motivation of this work is precisely not to explore exotic possibilities, such as extra dimensions, we will concentrate with the $f(R)$ theory in four spacetime dimensions, as the one describing gravitational interaction at the scale $\Lambda$. 

The paper is organized as follows: In \ref{fR_scalar} we will discuss the basic set up with $f(R)$ gravity and $n$ real scalar field. We will also present the perturbative expansion of the gravity plus matter action upto quadratic order. Subsequently, performing a path integral over the perturbations we will determine the effective action in \ref{eff_fR}. From the effective action we can read off the corrections to the mass of the particles and hence we can comment on breaking of conformal symmetry. For completeness, in \ref{equiv_str} we have discussed the equivalence of the above result involving $f(R)$ gravity with the scalar-tensor framework. Finally we conclude with a discussion on the results obtained.

\emph{Notations and Conventions:} We will work with mostly positive signature convention, such that the flat spacetime metric in Cartesian coordinates in four dimensional spacetime becomes $\textrm{diag}(-1,1,1,1)$. Lowercase Roman letters $a,b,\ldots$ denote spacetime indices and Uppercase Roman letters $A,B,\ldots$ count all the scalar fields in the problem. We also set the fundamental constants $c$ and $\hbar$ to unity. 

\section{f(R) gravity coupled with scalar fields: Perturbative expansion}\label{fR_scalar}

In this section, we will be studying the perturbations of both gravitational and scalar degrees of freedom for an interacting theory involving $f(R)$ gravity coupled with $n$ scalar fields. As emphasized in the introduction itself, in four spacetime dimensions, $f(R)$ gravity provides one of the most non-trivial higher curvature corrections to the Einstein-Hilbert action, free from Ostrogradsky's ghost. The final aim is to probe the mass renormalization of these scalar fields and determining the condition for vanishing of quadratic divergences in the one loop effective action for the $f(R)$ theory coupled with $n$ scalar fields. For this purpose, we will first express the action for the full system upto quadratic order in the gravitational as well as scalar field perturbation, whose subsequent integration over the gravitational and scalar perturbation will result into the one loop effective action\footnote{Certain aspects of one loop effective action for $f(R)$ gravity in the context of de Sitter background has been studied in \cite{Cognola:2005de,Bamba:2014mua,Bamba:2015uxa,Elizalde:2017mrn}. While for general ideas about one loop effective action, the reader may consult \cite{Buchbinder:1992rb}.}, which we describe in the next section. 

To begin with, we write down the gravitational plus matter action involving $f(R)$ gravity coupled with $n$ scalar field, which takes the following form,
\begin{align}\label{sbcs_action1}
\mathcal{A}=\frac{2}{\kappa^{2}}\int d^{4}x\sqrt{-g}f(R)+\int d^{4}x\sqrt{-g}\left\{-\frac{1}{2}g^{ab}\partial_{a}\Phi^{A}\partial_{b}\Phi_{A}-V(\Phi^{A}\Phi_{A}) \right\}~,
\end{align}
where $f(R)$ is an arbitrary function of the Ricci scalar, $\Phi_{A}=\delta _{AB}\Phi^{B}$ and $V(\Phi^{A}\Phi_{A})$ is an arbitrary function of the scalar fields, depending on the combination $\Phi^{A}\Phi_{A}$. Further, $\kappa^{2}=32\pi G$, where $G$ is the Newton's constant and $A$ takes values from $1,\ldots,n$ with repeated index denoting summation over all the scalar fields. Given the above action, we wish to expand the metric around the flat background and hence express the gravitational action upto quadratic order in the perturbation. Similarly, the scalar field will also be expanded around some background value and the matter action will involve both the background fields and perturbations upto quadratic order. For this purpose, we introduce the following perturbation for the metric as well as for the scalar field, which reads,
\begin{align}\label{sbcs_01}
g_{ab}=\eta_{ab}+\kappa h_{ab}~;\qquad \Phi^{A}=\phi^{A}+s^{A}~.
\end{align}
Here $\eta_{ab}$ is the background Minkowski spacetime and $\phi^{A}$ is the background value for the scalar field. Note that we will not consider the back-reaction of the scalar field on the background spacetime and hence $\phi^{A}$ can have non-zero value even though the background spacetime is flat. This is akin to the  notions of test fields living in the flat spacetime. Since we are perturbing around a flat background, which has vanishing Ricci scalar, it follows that we can also expand $f(R)$ around $R=0$. This yields, 
\begin{align}\label{sbcs_02}
f(R)=f(R=0)+f'(R=0)R+\frac{1}{2}f''(R=0)R^{2}+\frac{1}{3!}f'''(R=0)R^{3}+\mathcal{O}(R^{4})~.
\end{align}
Here, $f(R=0)$ acts as an effective cosmological constant term, which is not compatible with flat background and hence we must have $f(R=0)=0$. Note that this excludes the presence of any inverse powers of Ricci scalar in the theory. Further, the Ricci scalar, when expanded around the flat background depends on $h_{ab}$ linearly to the leading order \cite{Chakraborty:2020ktp}. Thus the term $R^{3}$ will start contributing to the action only at cubic order in $h_{ab}$. Hence, to the quadratic order, there will be no contribution in the gravitational action from the terms involving $R^{3}$ and higher powers of $R$. Therefore, for the purpose of this work, it will suffice if we restrict our attention upto the terms quadratic in the Ricci scalar in the expansion of $f(R)$ presented in \ref{sbcs_02}. Following this strategy, the gravitational Lagrangian upto quadratic order in $h_{ab}$, can be expressed in the following form \cite{Chakraborty:2020ktp},
\begin{align}\label{sbcs_03}
\mathcal{L}_{\rm grav}\equiv \frac{2}{\kappa^{2}}\sqrt{-g}f(R)&\simeq f'(R=0)\left(\frac{2}{\kappa^{2}}\sqrt{-g}R\right)+\frac{1}{2}f''(R=0)\left(\frac{2}{\kappa^{2}}\sqrt{-g}R^{2}\right)
\nonumber
\\
&=f'(R=0)\left(\frac{1}{2}h_{ab}\square h^{ab}-\frac{1}{2}h\square h-\partial_{a}h\partial_{b}h^{ab}+\partial_{a}h^{ab}\partial_{c}h^{c}_{b}\right)
\nonumber
\\
&+f''(R=0)\left(\partial_{a}\partial_{b}h^{ab}-\square h\right)^{2}~.
\end{align}
In order to arrive at the above expression we have not invoked any gauge choice, therefore this is an appropriate place to choose a particular gauge. It is customary to work in the Lorentz gauge, i.e., to impose the condition $\partial_{a}\{h^{ab}-(1/2)\eta^{ab}h\}=0$. We can either impose this gauge condition directly in \ref{sbcs_03} or, we can add a gauge fixing term to the gravitational Lagrangian. Such a gauge fixing term, in the present context, takes the following form,
\begin{align}\label{sbcs_04}
\mathcal{L}_{\rm gf}=\frac{1}{\xi}\left(\partial_{a}h^{ab}-\frac{1}{2}\partial^{b}h\right)^{2}+\frac{1}{\eta}\left(\partial_{a}\partial_{b}h^{ab}-\frac{1}{2}\square h\right)\left(\partial_{a}\partial_{b}h^{ab}-\frac{3}{2}\square h\right)~,
\end{align}
where $\xi$ and $\eta$ are constants to be fixed later. We see that under the Lorentz gauge condition, $\partial_{a}\{h^{ab}-(1/2)\eta^{ab}h\}=0$, the above gauge fixing Lagrangian identically vanishes, as it should. Thus the total Lagrangian density involving both gravitational as well as the gauge fixing term, when expanded upto quadratic order in the perturbation, will predominantly depend on the combinations $\{f'(R=0)+(1/\xi)\}$ as well as $\{f''(R=0)+(1/\eta)\}$ respectively. Thus to simplify the total Lagrangian we may as well choose the gauge fixing coefficients $\xi$ and $\eta$, such that, $\xi^{-1}=-f'(R=0)$ as well as, $\eta^{-1}=-f''(R=0)$. With these choices for $\xi$ and $\eta$ we obtain the total Lagrangian density of gravity and gauge fixing term together, yielding,
\begin{align}\label{grav_gf_fR}
\mathcal{L}_{\rm grav}+\mathcal{L}_{\rm gf}&=-\frac{f'(R=0)}{2}h_{ab}P^{ab;cd}\square h_{cd}+\frac{f''(R=0)}{4}\left(\square h\right)^{2}~,
\end{align}
where, $P^{ab;cd}\equiv (1/2)(\eta^{ab}\eta^{cd}-\eta^{ac}\eta^{bd}-\eta^{ad}\eta^{bc})$. Following the same line of arguments, as adopted for the gravitational perturbation above, it is possible to expand the matter Lagrangian containing terms upto quadratic order in $h_{ab}$ as well as in $s^{A}$. Such a decomposition has been performed in \ref{App_A}. We quote here the final result for the expansion of the matter Lagrangian along with the gravitational Lagrangian and the gauge fixing term, upto quadratic order in the perturbations, yielding,
\begin{align}
\mathcal{L}_{\rm tot}&\equiv\mathcal{L}_{\rm grav}+\mathcal{L}_{\rm gf}+\mathcal{L}_{\rm matter}
\nonumber
\\
&=-\frac{1}{2}h_{ab}P^{ab;cd}\left(f'(R=0)\square+\frac{\kappa^{2}}{2}V_{0}\right)h_{cd}-\frac{\kappa}{2}h\Big(s^{A}\partial_{A}V\Big)
\nonumber
\\
&\hskip 3 cm +\frac{1}{2}\delta_{AB}s^{A}\square s^{B}-\frac{1}{2}s^{A}s^{B}\partial_{A}\partial_{B}V+\frac{f''(R=0)}{4}h\square ^{2}h~,
\end{align}
where, $V_{0}$ is the scalar potential constructed out of the background scalar field $\phi^{A}$ in Minkowski spacetime. Also in arriving at the above expression we have assumed the background scalar fields $\phi^{A}$ to be constants. The above setup will be sufficient for describing the conformal symmetry through mass renormalization in the presence of higher curvature terms.   The above expression for the Lagrangian density $\mathcal{L}_{\rm tot}$ can be casted into a more suitable form, by using the explicit expression for $P^{ab;cd}$ as well as the following decomposition of the gravitational perturbation, $h_{ab}=H_{ab}+(1/4)\eta_{ab}h$, where $h$ is the trace of the gravitational perturbation and $H_{ab}$ denotes the traceless part. This results into,
\begin{align}\label{grav_decomp}
\mathcal{L}_{\rm tot}&=-\frac{1}{8}h\left(f'(R=0)\square+\frac{\kappa^{2}}{2}V_{0}-2f''(R=0)\square ^{2}\right)h+\frac{1}{2}H_{ab}\left(f'(R=0)\square+\frac{\kappa^{2}}{2}V_{0}\right)H^{ab}
\nonumber
\\
&\hskip 2cm -\frac{\kappa}{2}h\Big(\partial_{A}V\Big)s^{A}-\frac{1}{2}s^{A}\left(-\delta_{AB}\square +\partial_{A}\partial_{B}V\right)s^{B}~.
\end{align}
Note that the traceless part of the gravitational perturbation does not couple to the scalar fields, while the trace part couples with $s^{A}$. Moreover, the information about the background field is contained in the potential term $V_{0}$ and as we will see later, this term will be of prime importance in the mass renormalization scenario. Thus we may re-express the above Lagrangian in the following form,
\begin{align}\label{grav_decomp}
\mathcal{L}_{\rm tot}&=\frac{1}{2}H_{ab}\left(f'(R=0)\square+\frac{\kappa^{2}}{2}V_{0}\right)H^{ab}
\nonumber
\\
&-\frac{1}{2}
\begin{pmatrix}
\dfrac{h}{2} & s^{A}
\end{pmatrix}
\begin{pmatrix}
f'(R=0)\square+\frac{\kappa^{2}}{2}V_{0}-2f''(R=0)\square^{2} & \kappa\partial_{A}V\\
\\
\kappa\partial_{A}V  &-\delta_{AB}\square +\partial_{A}\partial_{B}V
\end{pmatrix}
\begin{pmatrix}
\dfrac{h}{2} \\
\\
s^{A}
\end{pmatrix}~.
\end{align}
This provides a natural division of the total Lagrangian density into two parts, one depending on the nine gravitational degrees of freedom encoded in the traceless tensor $H_{ab}$ and the other corresponds to $(n+1)$ degrees of freedoms. These involve the trace of the gravitational perturbation $h$ and $n$ scalar degrees of freedom $s^{A}$ respectively. As evident from \ref{grav_decomp}, we have a higher order differential operator $\square^{2}$ acting on the trace of the perturbation $h$, which has a coefficient proportional to $f''(R)$. Such higher derivative terms arise solely due to the presence of higher curvature corrections to the Einstein-Hilbert action. Interestingly, for theories with $f''(R=0)=0$, such higher derivative corrections will be absent and the Lagrangian, upto quadratic order in the perturbations will be identical to the general relativistic counterpart \cite{Meissner:2018mvq}. The above analysis provides us the final form for the Lagrangian involving the gravitational perturbation $h_{ab}$, the matter perturbations $s^{A}$, as well as the background field configurations $\phi^{A}$. In the next section, we will integrate out both the gravitational and the matter perturbations, thus determining the effective Lagrangian for the background scalar field $\phi^{A}$, which will be essential to comment on the issue of mass renormalization.
\section{Effective action and mass renormalization in f(R) gravity}\label{eff_fR}

In the previous section, we have derived the Lagrangian density for gravity and matter, where the gravitational sector is described by $f(R)$ gravity and the matter sector consists of $n$ real scalar fields. In the Lagrangian we have keept terms upto quadratic order in the metric as well as matter perturbations around flat Minkowski background, see \ref{grav_decomp}. The gravitational perturbation has been decomposed into its traceless part $H_{ab}$ and the trace $h$, which couples with the scalar perturbation $s^{A}$. Thus when we are computing the effective action for the background scalar field $\phi^{A}$ by integrating over the gravitational degrees of freedom and scalar perturbations, this interaction will affect the effective action for the scalar field. Furthermore, the trace part of the gravitational perturbation involves higher curvature terms (appearing as higher derivative operators) and hence these will affect the effective action for the background scalar field as well. Since the Lagrangian is quadratic in the perturbations, the path integral over the gravitational as well as scalar perturbations can be computed, leading to the following functional determinant,
\begin{align}
\mathcal{M}&=\left(f'(R=0)\square+\frac{\kappa^{2}}{2}V_{0}\right)^{-9/2}
\nonumber
\\
&\hskip 2 cm \times \textrm{det.}
\begin{pmatrix}
\left(f'(R=0)\square+\frac{\kappa^{2}}{2}V_{0}-2f''(R=0)\square ^{2}\right) & \frac{\kappa}{2}\left(\partial_{B}V\right) \\
\\
\frac{\kappa}{2}\left(\partial_{B}V\right) & -\delta_{AB}\square +\partial_{A}\partial_{B}V
\end{pmatrix}^{-1/2}~.
\end{align}
The first term comes from the nine traceless modes in $H_{ab}$, which are decoupled from the rest of the degrees of freedom. While the second factor, which is the determinant of a $(n+1)\times(n+1)$ matrix, comes from integrating out the trace part of $h$ and scalar perturbations $s^{A}$. Due to the coupling of the trace part $h$ with the scalar field perturbation $s^{A}$, the functional determinant involves cross terms. Thus we obtain, in the Fourier space, the following expression for the functional determinant $\mathcal{M}$,
\begin{align}\label{func_det_new}
\mathcal{M}&=\left(-f'(R=0)p^{2}+\frac{\kappa^{2}}{2}V_{0}\right)^{-9/2}
\nonumber
\\
&\hskip 2cm \times
\textrm{det.}
\begin{pmatrix}
\left(-f'(R=0)p^{2}+\frac{\kappa^{2}}{2}V_{0}-2f''(R=0)p^{4}\right) & \frac{\kappa}{2}\partial_{B}V \\
\\
\frac{\kappa}{2}\partial_{B}V & \delta_{AB}p^{2} +\partial_{A}\partial_{B}V
\end{pmatrix}^{-1/2}~,
\end{align}
where, we have used the result, $\square=-p^{2}$ in the Fourier space. Since we are adopting the mostly positive signature convention, we know that $p^{2}<0$ for causal fields. The first term in the functional determinant does not depend on the scalar field and hence the effect of the scalar field on the effective action is completely contained in the second term. This necessitates the evaluation of the second term, which is the determinant of a $(n+1)\times(n+1)$ matrix. Finding the determinant, can in principle be a humongous task, but to determine the quadratically divergent terms in the effective action, we simply need the sum of the eigenvalues of this functional determinant. The sum of eigenvalues can be found out, by simply expanding this determinant to first two leading orders in $-p^{2}$, which yields,
\begin{align}
(-1)^{n}~&\textrm{det.}
\begin{pmatrix}
\left(-f'(R=0)p^{2}+\frac{\kappa^{2}}{2}V-2f''(R=0)p^{4}\right) & \frac{\kappa}{2}\partial_{B}V \\
\\
\frac{\kappa}{2}\partial_{B}V & \delta_{AB}p^{2} +\partial_{A}\partial_{B}V
\end{pmatrix}
\nonumber
\\
&\hskip -1 cm =-2f''(R=0)\left(-p^{2}\right)^{n+2}+f'(R=0)\left(-p^{2}\right)^{n+1}+2f''(R=0)\left(\sum _{A=1}^{n}\partial_{A}\partial_{A}V\right)\left(-p^{2}\right)^{n+1}+\mathcal{O}\left(-p^{2}\right)^{n}
\nonumber
\\
&\hskip -1 cm =-2f''(R=0)\prod _{i=1}^{n+2}\left(-p^{2}-M_{i}^{2}\right)+\mathcal{O}(-p^{2})^{n}~,
\end{align}
where, the sum of the eigenvalues $M_{i}^{2}$, is determined by the coefficient of the $(-p^{2})^{n+1}$ term in the above expansion of the determinant, yielding,
\begin{align}\label{mass_fR}
\sum _{i=1}^{n+2}M_{i}^{2}=\frac{f'(R=0)}{2f''(R=0)}+\left(\sum _{A=1}^{n}\partial_{A}\partial_{A}V\right)~.
\end{align}
The above expression for the summation over all the eigenvalues is intimately connected with stability of the theory. Note that the positivity of the left hand side of \ref{mass_fR} demands, $f'(R=0)$ and $f''(R=0)$ to be positive, which is crucial for the stability of the $f(R)$ model under consideration. Further, we also require $\partial_{A}\partial_{A}V$ to be positive, which ensures that the background scalar field is near the minima of the scalar potential. This in turn ensures stability of the matter sector as well. Therefore we can conclude that the existence of such positive eigenvalues for the functional determinant is intimately connected with the stability of the $f(R)$ model and the scalar field Lagrangian respectively. Thus after all these algebraic manipulations, the functional determinant takes the following form,
\begin{align}\label{func_det_final}
\mathcal{M}\propto \left[2f''(R=0)\right]^{-1/2}\left(-f'(R=0)p^{2}+\frac{\kappa^{2}}{2}V_{0}\right)^{-9/2}\left[\prod _{i=1}^{n+2}\left(-p^{2}-M_{i}^{2}\right)\right]^{-1/2}~,
\end{align}
where some numerical factors have been neglected. It is worth emphasizing that, \ref{func_det_final} has no general relativity limit, since the analysis leading to \ref{func_det_final} demands $f''(R=0)\neq 0$. For the situation with $f''(R=0)=0$, we will have to go back to \ref{func_det_new}, which will reproduce the correct general relativity limit.

Having determined the functional determinant in the presence of higher curvature terms arising out of the path integral over the perturbations, let us work out the effective Lagrangian by wick rotating the background spacetime coordinates. This amounts to transforming to the Euclidean domain, yielding, $p^{2}=-p_{\rm E}^{2}$. Substituting the above transformation to the Euclidean domain in the functional determinant presented in \ref{func_det_final}, whose Logarithm yields the effective Lagrangian $\Gamma(\phi^{A})$ for the background scalar fields as,
\begin{align}
\Gamma(\phi^{A})&=\int _{0}^{\Lambda}\frac{d^{4}p_{\rm E}}{(2\pi)^{4}} \log \mathcal{M}
\nonumber
\\
&\hskip -1 cm =-\frac{1}{2}\int _{0}^{\Lambda}\frac{d^{4}p_{\rm E}}{(2\pi)^{4}}\left\{9\ln \left[\kappa^{2}\left(f'(R=0)p_{\rm E}^{2}+\frac{\kappa^{2}}{2}V_{0}\right)\right]+\sum _{i=1}^{n+2}\ln \left[\kappa^{2}\left(p_{\rm E}^{2}-M_{i}^{2}\right)\right]+\ln \left(\frac{2f''(R=0)}{\kappa^{2}}\right) \right\}
\nonumber
\\
&\hskip -1 cm =-\frac{1}{2}\int _{0}^{\Lambda}\frac{d^{4}p_{\rm E}}{(2\pi)^{4}}\Bigg[\left(n+11\right)\ln \left(\kappa^{2}p_{\rm E}^{2}\right)+9\ln \left(1+\frac{\kappa^{2}V_{0}}{2f'(R=0)p_{\rm E}^{2}} \right)+\sum _{i=1}^{n+2}\ln\left(1-\frac{M_{i}^{2}}{p_{\rm E}^{2}} \right) 
\nonumber
\\
&\hskip 2 cm +9\ln f'(R=0)+\ln \left(\frac{2f''(R=0)}{\kappa^{2}}\right) 
\Bigg]~.
\end{align}
Here $\Lambda$ is the cutoff scale of the problem, which is of the same order as the Planck scale $m_{\rm pl}\sim \kappa^{-1}$. The above integral involves three separate pieces, (a) integral over $\ln (\kappa^{2}p_{\rm E}^{2})$, (b) integral over $\ln \{1-(m^{2}/p_{\rm E}^{2})\}$ and (c) integral over constant pieces from the $f(R)$ gravity model. The first and third one, i.e., involving $\ln p_{\rm E}^{2}$ and constant terms when integrated over the four dimensional Euclidean manifold yields quartic divergences, 
\begin{align}
\int _{0}^{\Lambda}\frac{d^{4}p_{\rm E}}{(2\pi)^{4}}&\Big[(n+11)\ln \left(\kappa^{2}p_{\rm E}^{2}\right)+9\ln f'(R=0)+\ln \left(\frac{2f''(R=0)}{\kappa^{2}}\right)\Big]
\nonumber
\\
&=\frac{\Lambda^{4}}{32\pi^{2}}\left[(n+11)\ln \left(\kappa^{2}\Lambda^{2}\right)-\frac{(n+11)}{2}+9\ln f'(R=0)+\ln \left(\frac{2f''(R=0)}{\kappa^{2}}\right) \right]~.
\end{align}
Thus the quartically divergent term also depends on the presence of the higher curvature corrections, i.e., on the structure of the $f(R)$ Lagrangian. There are two possible ways to get rid of the quartically divergent term:
\begin{itemize}

\item  If we assume that near the Planck scale, some form of supersymmetry will be realized, then the Fermionic degrees of freedom will have a contribution to the quartically divergent term, which will be identical but of opposite sign compared to the Bosonic contribution above. Thus the quartically divergent term can be avoided. We must emphasize that the supersymmetry is employed at the \emph{Planck scale} and \emph{not} at any low energy scale.

\item In general relativity, the only way to get rid of the quartic divergent term is to employ Planck scale supersymmetry \cite{Meissner:2018mvq}, as described above. However, in the context of $f(R)$ gravity, the above divergent term can also be avoided by choosing $\ln (2f''(R=0)/\kappa^{2})+9\ln f'(R=0)=-(n+11)\ln (\kappa^{2}\Lambda^{2})+(1/2)(n+11)$, \emph{without} requiring supersymmetry at all. In particular, for the Starobinsky model, $f(R)=R+\alpha R^{2}$ the above condition yields, $\ln (4\alpha m_{\rm Pl}^{2})=-(n+11)\ln (\Lambda^{2}/m_{\rm Pl}^{2})+(1/2)(n+11)$. Thus for $\Lambda <m_{\rm Pl}$, we can choose $\alpha m_{\rm Pl}^{2}>(1/4)$, while for $\Lambda=m_{\rm Pl}$, $\alpha m_{\rm Pl}^{2}$ is uniquely determined by the number of scalar field species in the problem. This suggests that the dimensionless coupling parameter $\alpha m_{\rm Pl}^{2}$ acts as a natural cutoff scale for the theory, whose appropriate choice will cancel the quartically divergent term. Therefore, use of the higher curvature terms may allow one to set the quartic divergent terms to zero, \emph{without} invoking supersymmetry and may give an idea about the coupling parameters appearing in the $f(R)$ model. 

\end{itemize}
On the other hand, the integral of $\ln [1-(m^{2}/p_{\rm E}^{2})]$ over the four dimensional Euclidean manifold has both quadratically divergent as well as Logarithmically divergent term. The contribution to the quadratically divergent term from the effective action can be expressed in the following form, 
\begin{align}
\Gamma^{\rm quad}(\phi^{A})=\frac{\Lambda^{2}}{32\pi^{2}}\left\{\frac{f'(R=0)}{2f''(R=0)}+\left(\sum _{A}\partial_{A}\partial_{A}V\right)-\frac{9}{2}\kappa^{2}V_{0}\right\}+\mathcal{O}(\ln \Lambda)~.
\end{align}
Thus we observe that alike the quartically divergent term, the quadratically divergent contribution to the effective action also depends heavily on the presence of the parameter $f''(R=0)$. Further the stability of the $f(R)$ theory demands $f''(R)$ as well as $f'(R)$ to be positive, which in turn leads to a positive contribution to the quadratically divergent term of the effective action. Using the following generic form for the potential, 
\begin{align}\label{gen_pot}
V(\phi^{A})=\frac{1}{2}m_{AB}^{2}\phi^{A}\phi^{B}+\frac{1}{4!}\lambda_{ABCD}\phi^{A}\phi^{B}\phi^{C}\phi^{D}+\mathcal{O}(\kappa^{2})~,
\end{align}
where $m_{AB}^{2}$ is the mass matrix of the $n$ scalar fields and $\lambda_{ABCD}$ are the dimensionless coupling constants of quartic interaction between the scalar fields, the quadratically divergent part of the effective action reads,
\begin{align}
\Gamma^{\rm quad}(\phi^{A})=\frac{\Lambda^{2}}{32\pi^{2}}\left\{\frac{f'(R=0)}{2f''(R=0)}+\left(\sum _{A=1}^{n}m_{AA}^{2}\right)
+\left(\frac{1}{2}\sum _{A=1}^{n}\lambda_{AACD}-\frac{9}{4}\kappa^{2}m_{AB}^{2}\right)\phi^{A}\phi^{B}+\mathcal{O}(\phi^{4})\right\}~.
\end{align}
Given the above expression for the quadratically divergent part of the effective action, one can read off the corrections to the mass matrix $m_{AB}^{2}$ as the coefficient of quadratic terms in the background scalar field, which takes the form
\begin{align}
\delta m_{AB}^{2}=-\frac{\Lambda^{2}}{16\pi^{2}}\left(\frac{1}{2}\sum _{A=1}^{n}\lambda_{AACD}-\frac{9}{4}\kappa^{2}m_{AB}^{2}\right)~.
\end{align}
It is evident that, in the absence of gravitational interaction, the second term in the above expression will be absent, since it explicitly depends on the gravitational constant. If initially the masses of the scalar fields were small enough, i.e., $m_{AB}^{2}\ll \Lambda^{2}$, then for $\kappa\sim m_{\rm Pl}^{-1}\sim \Lambda^{-1}$, it follows that $\kappa^{2}m_{AB}^{2}\ll \mathcal{O}(1)$. Therefore, the corrections to the mass matrix due to quadratic divergences in the one loop effective action have negligible contributions from the gravitational corrections, be it Einstein gravity or higher curvature theory. Hence the renormalized mass matrix will depend solely on the bare couplings present in the theory and can be set to zero by choosing the scale $\Lambda$ and bare coupling parameters appropriately (for a similar scenario in the context of Higg's Boson, see \cite{Hamada:2012bp}). Then we have $\delta m_{AB}^{2}\sim \mathcal{O}(\ln \Lambda)$ and hence the masses of the scalar fields can be consistently kept small all the way upto Planck scale. Therefore, the conformal invariance of the theory will only be softly broken.    

\section{Equivalence of the effective action and mass renormalization in the scalar-tensor representation}\label{equiv_str}

In the previous sections, we have explicitly demonstrated how the presence of higher curvature terms in the guise of $f(R)$ gravity affects the effective action \emph{but} still keeps the corrections to the counter term in the mass renormalization negligible. We have also demonstrated that the stability of the $f(R)$ theory is intimately connected with the existence of a well-defined effective action for the gravity plus scalar system. However, we also know that any $f(R)$ Lagrangian can equivalently be expressed in the scalar-tensor representation as well \cite{Sotiriou:2006hs,Briscese:2006xu,Chakraborty:2016ydo}. Thus the above conclusions should hold true in the scalar-tensor representation as well. This is what we will explicitly establish in this section. For this purpose, it is instructive to start with the standard $f(R)$ Lagrangian along with the matter sector involving $n$ scalar fields, and from which we will make a transition to the Einstein frame. This procedure involves three steps. First of all, one rewrites the original Lagrangian for $f(R)$ gravity in the following form, 
\begin{align}
\mathcal{L}_{\rm Jordan}=\frac{2}{\kappa^{2}}\sqrt{-g}\left[Rf'(\chi)-\left\{\chi f'(\chi)-f(\chi)\right\} \right]+\sqrt{-g}L_{\rm matter}~,
\nonumber
\end{align}
where $\chi$ is an auxiliary field. Note that the variation of the above Lagrangian density with respect to the auxiliary field $\chi$ yields, the equation of motion of $\chi$ to be, $R=\chi$. Then the on-shell value of the above Lagrangian becomes identical to the Lagrangian for $f(R)$ gravity coupled with matter field. At the second step one uses the conformal transformation, $\bar{g}_{ab}=\Omega^{2}g_{ab}$. Under such a conformal transformation the Ricci scalar gets modified, such that, $R=\Omega^{2}\bar{R}-6\bar{g}^{ab}\nabla_{a}\Omega \nabla_{b}\Omega+6\Omega^{2}\bar{\square}\ln \Omega$. Therefore, the above Lagrangian in terms of the conformally transformed metric $\bar{g}_{ab}$ becomes,
\begin{align}
\mathcal{L}_{\rm Jordan}&=\frac{2}{\kappa^{2}}\Omega^{-4}\sqrt{-\bar{g}}\left[f'(\chi)\left\{\Omega^{2}\bar{R}-6\bar{g}^{ab}\nabla_{a}\Omega \nabla_{b}\Omega+6\Omega^{2}\bar{\square}\ln \Omega \right\}-\left\{\chi f'(\chi)-f(\chi)\right\} \right]+\Omega^{-4}\sqrt{-\bar{g}}\bar{L}_{\rm matter}
\nonumber
\\
&=\frac{2}{\kappa^{2}}\sqrt{-\bar{g}}\left\{\Omega^{-2}f'(\chi)\right\}\left(\bar{R}-6\bar{g}^{ab}\nabla_{a}\ln \Omega \nabla_{b}\ln \Omega+6\bar{\square}\ln \Omega\right)
\nonumber
\\
&\hskip 2 cm -\frac{2}{\kappa^{2}}\sqrt{-\bar{g}}\Omega^{-4}\left\{\chi f'(\chi)-f(\chi)\right\}+\Omega^{-4}\sqrt{-\bar{g}}\bar{L}_{\rm matter}~.
\end{align}
The third and last step involves relating $\Omega$ to the auxiliary field $\chi$ and introduce a scalar field $\psi$, such that $\Omega^{2}=f'(\chi)$ and $\kappa \psi=2\sqrt{6}\ln \Omega$. With these identifications, the gravitational Lagrangian takes the following form in the Einstein frame,
\begin{align}
\mathcal{L}_{\rm E}&=\frac{2}{\kappa^{2}}\sqrt{-\bar{g}}\bar{R}-\frac{1}{2}\sqrt{-\bar{g}}\bar{g}^{ab}\bar{\nabla}_{a}\psi \bar{\nabla}_{b}\psi+\frac{12}{\kappa^{2}}\sqrt{-\bar{g}}\bar{\square}\ln \Omega
\nonumber
\\
&\hskip 2 cm -\frac{2}{\kappa^{2}}\sqrt{-\bar{g}}\Omega^{-4}\left\{\chi f'(\chi)-f(\chi)\right\}+\Omega^{-4}\sqrt{-\bar{g}}L_{\rm matter}~.
\end{align}
In the above expression for the Lagrangian, the term depending on $\bar{\square}\ln \Omega$ will not contribute, since it will yield a boundary term when integrated over four dimensional spacetime with conformally transformed metric $\bar{g}_{ab}$. Thus neglecting such boundary contributions we obtain the Lagrangian of the gravity plus matter system in the conformally transformed frame, to yield, 
\begin{align}\label{Lagrangian_Einstein}
\mathcal{L}_{\rm E}&=\frac{2}{\kappa^{2}}\sqrt{-\bar{g}}\bar{R}+\sqrt{-\bar{g}}\left\{-\frac{1}{2}\bar{g}^{ab}\bar{\nabla}_{a}\psi \bar{\nabla}_{b}\psi -W(\psi) \right\}
\nonumber
\\
&\hskip 2 cm +\sqrt{-\bar{g}}\left\{-\frac{1}{2}\Omega^{-2}\delta_{AB}\bar{g}^{ab}\partial_{a}\Phi^{A}\partial_{b}\Phi^{B}-\Omega^{-4}V(\Phi) \right\}~.
\end{align}
Here we have introduced the quantity $W(\psi)$, which can be defined as,
\begin{align}\label{scalar_pot}
W(\psi)\equiv \frac{2}{\kappa^{2}}\Omega^{-4}\left\{\chi f'(\chi)-f(\chi)\right\};\qquad \kappa \psi=\sqrt{6}\ln f'(\chi)~.
\end{align}
The Lagrangian in the Einstein frame, as depicted in \ref{Lagrangian_Einstein} must be contrasted with the Lagrangian presented in \cite{Meissner:2018mvq}. There are two main differences between the two Lagrangians --- (a) The kinetic term for the scalar field $\Phi^{A}$ in the Lagrangian of \ref{Lagrangian_Einstein} is \emph{not} canonical, as it couples to the scalar degree of freedom arising from the $f(R)$ model. While in \cite{Meissner:2018mvq}, the kinetic terms for the scalar fields are strictly canonical; (b) The potential term is \emph{no longer} simply $V(\Phi_{A}\Phi^{A})$, rather it is coupled with a function of the scalar degree of freedom from the $f(R)$ gravity model, which is also different from \cite{Meissner:2018mvq}. If we consider the limit $f(R)\rightarrow R$, then we will have $\Omega \rightarrow 1$ and hence the above Lagrangian will indeed reduce to that of \cite{Meissner:2018mvq}. This can be taken to be a consistency check of the computation presented here.

In what follows we will resort to the same strategy as in the previous section, i.e., expand the above Lagrangian around flat spacetime, such that $\bar{g}_{ab}=\eta_{ab}+h_{ab}$. If we want a correspondence with the $f(R)$ gravity, then for the background spacetime we must have $\Omega=1$ and hence $\psi=0$ for the background spacetime. Thus in the above Lagrangian $\psi$ itself can be considered as a perturbation about flat background. Therefore, in the scalar-tensor representation we have three perturbation variables, the gravitational perturbation $h_{ab}$, scalar perturbation $\psi$ and matter perturbation $s^{A}$. We will now expand the gravity plus matter Lagrangian to quadratic order in these perturbation variables and integrate over them in order to derive the effective action.  

In order to proceed further, we will borrow the results from the previous sections and hence determine the Ricci scalar upto quadratic order in the gravitational perturbation $h_{ab}$. This can be achieved by setting $f'(R=0)=1$ and $f''(R=0)=0$ in \ref{grav_gf_fR}, such that the Einstein-Hilbert term in the Lagrangian, along with the gauge fixing term yields,
\begin{align}
\mathcal{L}_{\rm gr}+\mathcal{L}_{\rm gf}&=\frac{1}{2}h_{ab}\square h^{ab}-\frac{1}{4}h\square h
=-\frac{1}{2}h_{ab}P^{ab;cd}\square h_{cd};
\qquad P^{ab;cd}\equiv \frac{1}{2}\left(\eta^{ab}\eta^{cd}-\eta^{ac}\eta^{bd}-\eta^{ad}\eta^{bc}\right)~.
\end{align}
The matter Lagrangian constructed out of the scalar $\psi$ originating from the scalar-tensor representation of the $f(R)$ gravity, when expanded to quadratic order in the perturbation variables can be expressed as,
\begin{align}
\mathcal{L}_{\rm matter}^{\rm scalar-tensor}&=\sqrt{-\bar{g}}\left\{-\frac{1}{2}\bar{g}^{ab}\bar{\nabla}_{a}\psi \bar{\nabla}_{b}\psi -W(\psi) \right\}
\nonumber
\\
&=-\frac{\kappa}{2}h\psi \left(\partial W/\partial \psi \right)-\frac{1}{2}\eta^{ab}\partial_{a}\psi \partial_{b}\psi
-\frac{1}{2}\psi^{2}\left(\partial^{2}W/\partial \psi^{2}\right)~,
\end{align}
where the potential term $W(\psi)$ has been expanded as a Taylor series around the background $\psi=0$. Finally the original matter Lagrangian involving $n$ scalar fields can also be expanded upto quadratic order in the perturbations, which has been performed in \ref{App_B}. Therefore the total Lagrangian involving all the perturbations at the quadratic order takes the following form, 
\begin{align}
\mathcal{L}_{\rm gr}&+\mathcal{L}_{\rm gf}+\mathcal{L}_{\rm matter}^{\rm scalar-tensor}+\mathcal{L}_{\rm matter}
\nonumber
\\
&=\frac{1}{2}H_{ab}\left(\square+\frac{\kappa^{2}}{2}V_{0}\right) H^{ab}-\frac{1}{8}h\left(\square+\frac{\kappa^{2}}{2}V_{0}\right)h-\frac{\kappa}{2}h\psi \left\{\left(\partial W/\partial \psi \right)-\frac{2\kappa V_{0}}{\sqrt{6}} \right\}-\frac{1}{2}\psi^{2}\left\{\left(\partial^{2}W/\partial \psi^{2}\right)+\frac{2\kappa^{2}V_{0}}{3} \right\}
\nonumber
\\
&-\frac{1}{2}\eta^{ab}\partial_{a}\psi \partial_{b}\psi-\frac{\kappa}{2}hs^{A}\partial_{A}V-\frac{1}{2}\delta_{AB}\eta^{ab}\partial_{a}s^{A}\partial_{b}s^{B}-\frac{1}{2}s^{A}s^{B}\partial_{A}\partial_{B}V
-2\left(-\frac{\kappa \psi}{\sqrt{6}}\right)s^{A}\partial_{A}V~.
\end{align}
Here $H_{ab}$ is the traceless part of the gravitational perturbation and $h$ is the trace part. It should also be emphasized that the perturbation $\psi$ in the scalar-tensor sector is actually originating from the higher curvature corrections present in the $f(R)$ theory of gravity. Since our interest lies in the determination of the effective action for the background scalar field, we need to integrate over all the perturbed quantities, $H_{ab}$, $h$, $\psi$ and $s^{A}$. Such a functional integral over all the perturbed quantities yield the following functional determinant,
\begin{align}
\mathcal{M}&=\left(\square+\frac{\kappa^{2}}{2}V_{0}\right)^{-9/2}
\nonumber
\\
&\times \textrm{det.}
\begin{pmatrix}
\square +\frac{\kappa^{2}}{2}V_{0}& \frac{\kappa}{2} \left\{\left(\partial W/\partial \psi \right)-\frac{2\kappa V_{0}}{\sqrt{6}} \right\} & \frac{\kappa}{2}\partial_{B}V 
\\
\frac{\kappa}{2} \left\{\left(\partial W/\partial \psi \right)-\frac{2\kappa V_{0}}{\sqrt{6}} \right\} & -\square +\left\{\left(\partial^{2}W/\partial \psi^{2}\right)+\frac{2\kappa^{2}V_{0}}{3} \right\} & -\left(\frac{2\kappa}{\sqrt{6}}\right)\partial_{A}V
\\
\frac{\kappa}{2}\partial_{B}V & -\left(\frac{2\kappa}{\sqrt{6}}\right)\partial_{A}V & -\delta_{AB}\square +\partial_{A}\partial_{B}V
\end{pmatrix}^{-1/2}~.
\end{align}
Note that in the case of $f(R)$ gravity, the $n$ scalar fields were coupled with the trace of the gravitational perturbation. In the present context, along with the trace part, the scalar fields are also coupled to $\psi$, the field appearing from the transition of $f(R)$ theory to scalar-tensor representation. It is instructive to transform the above functional determinant to the Fourier space, which amounts to transforming $\square$ to $-p^{2}$ in the above expression. Therefore, the functional determinant in the Fourier space becomes,  
\begin{align}
\mathcal{M}&=\left(-p^{2}+\frac{\kappa^{2}}{2}V_{0}\right)^{-9/2}
\nonumber
\\
&\times \textrm{det.}
\left(\begin{array}{ccc}
-p^{2} +\frac{\kappa^{2}}{2}V_{0}& \frac{\kappa}{2} \left\{\left(\partial W/\partial \psi \right)-\frac{2\kappa V_{0}}{\sqrt{6}} \right\} & \frac{\kappa}{2}\partial_{B}V 
\\
\frac{\kappa}{2} \left\{\left(\partial W/\partial \psi \right)-\frac{2\kappa V_{0}}{\sqrt{6}} \right\} & p^{2} +\left\{\left(\partial^{2}W/\partial \psi^{2}\right)+\frac{2\kappa^{2}V_{0}}{3} \right\} & -\left(\frac{2\kappa}{\sqrt{6}}\right)\partial_{A}V
\\
\frac{\kappa}{2}\partial_{B}V & -\left(\frac{2\kappa}{\sqrt{6}}\right)\partial_{A}V & \delta_{AB}p^{2} +\partial_{A}\partial_{B}V
\end{array}\right)^{-1/2}~.
\end{align}
In order to find out the functional determinant one needs to work out the determinant of the $(n+2)\times (n+2)$ matrix originating from the $n$ scalar fields, the trace of the gravitational perturbation $h$ and the additional scalar field $\psi$. This in practice is a very complicated computation to perform, but for our purpose of determining the leading order divergent contributions in the effective action, it will suffice to consider the first two leading order powers of the momentum. This yields,
\begin{align}
\left(-1\right)^{n+1}&\textrm{det.}
\begin{pmatrix}
-p^{2}+\frac{\kappa^{2}}{2}V_{0}& \frac{\kappa}{2} \left\{\left(\partial W/\partial \psi \right)-\frac{2\kappa V_{0}}{\sqrt{6}} \right\} & \frac{\kappa}{2}\partial_{B}V 
\\
\frac{\kappa}{2} \left\{\left(\partial W/\partial \psi \right)-\frac{2\kappa V_{0}}{\sqrt{6}} \right\} & p^{2} +\left\{\left(\partial^{2}W/\partial \psi^{2}\right)+\frac{2\kappa^{2}V_{0}}{3} \right\} & -\left(\frac{2\kappa}{\sqrt{6}}\right)\partial_{A}V
\\
\frac{\kappa}{2}\partial_{B}V & -\left(\frac{2\kappa}{\sqrt{6}}\right)\partial_{A}V & \delta_{AB}p^{2} +\partial_{A}\partial_{B}V
\end{pmatrix}
\nonumber
\\
&\hskip -1 cm \simeq \left(-p^{2} +\frac{\kappa^{2}}{2}V_{0}\right)\left[-p^{2}-\left\{\left(\partial^{2}W/\partial \psi^{2}\right)+\frac{2\kappa^{2}V_{0}}{3} \right\}\right]\left[\left(-p^{2}\right)^{n}-\left(-p^{2}\right)^{n-1}\sum_{A=1}^{n}\partial_{A}\partial_{A}V\right]
\nonumber
\\
&=\left(-p^{2}\right)^{n+2}-\left(-p^{2}\right)^{n+1}\left\{-\frac{\kappa^{2}}{2}V_{0}+\left(\partial^{2}W/\partial \psi^{2}\right)+\frac{2\kappa^{2}V_{0}}{3}+\sum_{A}\partial_{A}\partial_{A}V \right\}
\nonumber
\\
&=\prod _{i=1}^{n+2}\left(-p^{2}-M_{i}^{2}\right)~,
\end{align}
where, $M_{i}^{2}$ are certain characteristic mass scales associated with this problem, satisfying the following result, 
\begin{align}
\sum _{i}M_{i}^{2}=\left(\partial^{2}W/\partial \psi^{2}\right)+\sum_{A}\partial_{A}\partial_{A}V+\frac{\kappa^{2}V_{0}}{6}~.
\end{align}
Again the left hand side of the above equation must be positive definite, which requires the potential $W(\psi)$ as well as the potential $V(\phi)$ to have a minimum. This shows another crucial difference with \cite{Meissner:2018mvq}, as in the present context the positive definiteness of the eigenvalues $M_{i}^{2}$ not only requires the potential $V(\Phi^{A}\Phi_{A})$ to have a minima, but also it demands existence of minima for $W(\psi)$ as well. 

Thus once again the stability of the theory is intimately connected with the positivity of $M_{i}^{2}$, which is extremely important for (almost) conformal invariance of the theory. Hence the equivalence between the stability of the $f(R)$ theory and its scalar-tensor representation is manifest from the above analysis. 

Let us proceed further and determine the effective action by taking the Logarithm of the functional determinant presented above by integrating over the four momentum. It is advantageous to translate the above results into Euclidean manifold by performing a Wick rotation. Under such a transformation, $p^{2}\rightarrow -p_{\rm E}^{2}$ and hence the effective Lagrangian in the Euclidean domain will read,
\begin{align}
\Gamma(\phi^{A})&=\int _{0}^{\Lambda}\frac{d^{4}p_{\rm E}}{(2\pi)^{4}} \log \mathcal{M}
\nonumber
\\
&\hskip -1 cm =-\frac{1}{2}\int _{0}^{\Lambda}\frac{d^{4}p_{\rm E}}{(2\pi)^{4}}\left\{9\ln \left[\kappa^{2}\left(p_{\rm E}^{2}+\frac{\kappa^{2}}{2}V_{0}\right)\right]+\sum _{i=1}^{n+2}\ln \left[\kappa^{2}\left(p_{\rm E}^{2}-M_{i}^{2}\right)\right]\right\}
\nonumber
\\
&\hskip -1 cm =-\frac{1}{2}\int _{0}^{\Lambda}\frac{d^{4}p_{\rm E}}{(2\pi)^{4}}\Bigg[\left(n+11\right)\ln \left(\kappa^{2}p_{\rm E}^{2}\right)+9\ln \left(1+\frac{\kappa^{2}V_{0}}{2p_{\rm E}^{2}} \right)+\sum _{i=1}^{n+2}\ln\left(1-\frac{M_{i}^{2}}{p_{\rm E}^{2}} \right) \Bigg]~.
\end{align}
Here also, the integral has two main ingredients --- (a) terms involving $\ln(\kappa^{2}p_{\rm E}^{2})$ and (b) terms involving $\ln[1-(m^{2}/p_{\rm E}^{2})]$. The integral over $\ln(\kappa^{2}p_{\rm E}^{2})$ will lead to quartically divergent contribution, which can be set to zero by assuming existence of supersymmetry at a high energy scale. This is because, existence of supersymmetry will induce an identical but opposite contribution coming from the Fermionic sector as well, which will make the total contribution of Bosonic and Fermionic system to be vanishing. On the other hand the integral involving $\ln[1-(m^{2}/p_{\rm E}^{2})]$ will yield a quadratically divergent contribution along with a Logarithmic correction term, such that the effective action to leading order becomes,
\begin{align}
\Gamma^{\rm quad}(\phi^{A})=\frac{\Lambda^{2}}{32\pi^{2}}\Big[\left(\partial^{2}W/\partial \psi^{2}\right)+\sum_{A=1}^{n}\partial_{A}\partial_{A}V+\frac{\kappa^{2}V_{0}}{6}-\frac{9}{2}\kappa^{2}V_{0}\Big]~.
\end{align}
Simplifying further and using the generic form for the scalar potential as presented in \ref{gen_pot}, along with the expression for $(d^{2}W/d\psi^{2})$ from \ref{App_C} we obtain the following expression for the effective action,
\begin{align}
\Gamma^{\rm quad}(\phi^{A})=\frac{\Lambda^{2}}{32\pi^{2}}\Bigg[\frac{1}{3}\frac{f'(R=0)^{2}}{f''(R=0)}+\sum_{A=1}^{n}m_{AA}^{2}+\left(\frac{1}{2}\sum_{A=1}^{n}\lambda_{AACD}-\frac{13}{6}\kappa^{2}m_{CD}^{2}\right)\phi^{C}\phi^{D}+\mathcal{O}(\phi^{4})\Bigg]~.
\end{align}
It is evident from the above expression that except for some numerical factors of $\mathcal{O}(1)$, the quadratically divergent piece of the effective action in the scalar-tensor representation is identical to the one in $f(R)$ theory. This explicitly demonstrates the equivalence between the two. Further, the conclusion regarding smallness of the quadratically divergent counter term in mass renormalization also remain unchanged. To see this explicitly, we write down the corrections $\delta m_{AB}^{2}$ to the mass matrix $m_{AB}^{2}$, below
\begin{align}
\delta m_{CD}^{2}=-\frac{\Lambda^{2}}{16\pi^{2}}\left(\frac{1}{2}\sum_{A=1}^{n}\lambda_{AACD}-\frac{13}{6}\kappa^{2}m_{CD}^{2}\right)~.
\end{align}
As evident from the above expression, if the elements of the mass matrix $m_{AB}^{2}$ were much smaller than the scale $\Lambda$, we have $\kappa^{2}m_{AB}^{2}\ll \mathcal{O}(1)$ as well. Therefore, the corrections to the mass matrix arising out of the gravitational interaction are negligible. Hence the masses will remain smaller even when the gravitational and higher loop effects are taken into account. This suggests that the conformal symmetry of the original Lagrangian will remain weakly broken, as desired. Note that we had arrived at the same conclusion in the context of $f(R)$ theory as well. Furthermore, if $\kappa^{2}m_{AB}^{2}\ll \mathcal{O}(1)$, and the potential due to $\psi$ dominates at high energy, it follows that the quadratically divergent term can be made smaller altogether. 

This analysis serves two purposes for us. Firstly, it strengthens the equivalence between the $f(R)$ theory and its scalar-tensor representation in the context of one loop effective action and mass renormalization. Secondly, it demonstrates the robustness of the fact that gravitational interaction has very little effect on the mass renormalization of the matter fields, provided the original masses were small (compared to the scale $\Lambda$) to begin with. Moreover this is true even in the context of higher curvature gravity. 
\section{Concluding Remarks}

The gauge hierarchy problem and its possible resolution has taken the centerstage of the high energy physics for the last decade. Even after advocating several intriguing and exotic possibilities to bypass the gauge hierarchy problem, none has been realized so far in the experiments. This has provided significant motivation to look for other alternative scenarios, where the gauge hierarchy problem can be addressed without deviating much from the Standard Model. One such possibility is the idea of softly broken conformal invariance, where the bare couplings of the theory are chosen in such a manner that counter term to the mass renormalization is unaffected by quadratically divergent contributions arising out of higher loop corrections. Since gravitational interaction is universal it will necessarily couple with the matter fields, thereby modifying the scenario presented above. As the scale at which bare couplings are evaluated is $\mathcal{O}(m_{\rm Pl})$, it is expected that the gravitational interaction will inherit higher curvature corrections. Following which, we have discussed the effect of such higher curvature corrections, in the form of $f(R)$ gravity and its implications for mass renormalization scenario. 

Our analysis makes it clear that gravitational interactions, be it Einstein gravity or f(R) gravity, has very little effect on the mass renormalization scenario, provided the bare masses of the scalar fields in the Lagrangian were small compared to the cut off scale $\Lambda$ ($\sim \mathcal{O}(m_{\rm Pl})$) to begin with. Therefore, the quadratically divergent piece in the counter term becomes identical to the contribution from flat spacetime and can be set to zero by choosing the bare couplings appropriately. Thus the masses of the scalar fields can be kept identical to the original values $m^{2}$, satisfying $m^{2}\ll \Lambda^{2}$. This suggests that the conformal symmetry will be approximately preserved at the energy scale $\Lambda$. 

Even though the higher curvature corrections do not affect the mass renormalization scenario directly, it does have indirect consequences. First of all, the functional determinant, crucial in finding out the one loop effective action, will have positive eigenvalues if and only if the $f(R)$ theory is stable, i.e., free from any ghost modes. Secondly, the quartically divergent term in the effective action can be eliminated without any necessity to invoke supersymmetry, but by choosing the couplings in the $f(R)$ model in a suitable manner. Finally, we have also demonstrated that all these results mentioned above hold correct in the scalar-tensor representation as well. This shows another instance of equivalence between $f(R)$ gravity with its scalar-tensor representation. In a nutshell, following the analysis of this work we can safely conclude that weakly (or, softly) broken conformal invariance for Standard Model seems to be a viable candidate to address the gauge hierarchy problem and is minimally affected by the gravitational interactions, described by either general relativity or higher curvature corrections transcending general relativity.

\section*{Acknowledgements}

The author (S.C.) thanks Hermann Nicolai for useful discussions and suggestions at various stages of this work and acknowledges Max-Planck Society for awarding Max-Planck-India Mobility Grant. S.C. also thanks AEI, Potsdam for warm hospitality and financial support, where initial stages of this work were carried out. The research of S.C. is supported in part by the INSPIRE Faculty fellowship from Department of Science and Technology, Government of India (Reg. No. DST/INSPIRE/04/2018/000893).
\appendix
\labelformat{section}{Appendix #1} 
\labelformat{subsection}{Appendix #1}
\section*{Appendices}
\section{Expansion of the matter Lagrangian upto quadratic order in the perturbations}\label{App_A}

The matter Lagrangian involving $n$ real scalar fields, minimally coupled with gravity can also be expanded upto quadratic order in the matter perturbation $s^{A}$ and gravitational perturbation $h_{ab}$. The computation of the action expanded upto second order, can be performed along the following lines, 
\begin{align}
\mathcal{L}_{\rm matter}&=\sqrt{-g}\left\{\delta_{AB}\left(-\frac{1}{2}g^{ab}\partial_{a}\Phi^{A}\partial_{b}\Phi^{B}\right)-V(\Phi)\right\}
\nonumber
\\
&=\left[1+\frac{\kappa}{2}h-\frac{\kappa^{2}}{4}\left(h^{\alpha \beta}h_{\alpha \beta}-\frac{1}{2}h^{2}\right)\right]\Bigg\{-\frac{1}{2}\delta_{AB}\left[\left(\eta^{ab}-\kappa h^{ab}+\kappa^{2}h^{ac}h^{b}_{c}\right)\partial_{a}\left(\phi^{A}+s^{A}\right)\partial_{b}\left(\phi^{B}+s^{B}\right)\right]
\nonumber
\\
&\hskip 2 cm -V(\phi)-s^{A}\partial_{A}V-\frac{1}{2}s^{A}s^{B}\partial_{A}\partial_{B}V\Bigg\}
\nonumber
\\
&=\left[1+\frac{\kappa}{2}h-\frac{\kappa^{2}}{4}\left(h^{\alpha \beta}h_{\alpha \beta}-\frac{1}{2}h^{2}\right)\right]\Bigg\{-\frac{1}{2}\delta_{AB}\Big[\left(\eta^{ab}-\kappa h^{ab}+\kappa^{2}h^{ac}h^{b}_{c}\right)
\nonumber
\\
&\hskip 2 cm \times\left(\partial_{a}\phi^{A}\partial_{b}\phi^{B}+2\partial_{a}\phi^{A}\partial_{b}s^{B}+\partial_{a}s^{A}\partial_{b}s^{B}\right)\Big] -V(\phi)-s^{A}\partial_{A}V-\frac{1}{2}s^{A}s^{B}\partial_{A}\partial_{B}V\Bigg\}
\nonumber
\\
&=\left[1+\frac{\kappa}{2}h-\frac{\kappa^{2}}{4}\left(h^{\alpha \beta}h_{\alpha \beta}-\frac{1}{2}h^{2}\right)\right]\Bigg[-\frac{1}{2}\delta_{AB}\left(\eta^{ab}\partial_{a}\phi^{A}\partial_{b}\phi^{B}\right)-V(\phi)
\nonumber
\\
&\hskip 1 cm -\frac{1}{2}\delta_{AB}\left(-\kappa h^{ab}+\kappa^{2}h^{ac}h^{b}_{c}\right)\partial_{a}\phi^{A}\partial_{b}\phi^{B}-\delta_{AB}\eta^{ab}\partial_{a}\phi^{A}\partial_{b}s^{B}-s^{A}\partial_{A}V+\delta_{AB}\kappa h^{ab}\partial_{a}\phi^{A}\partial_{b}s^{B}
\nonumber
\\
&\hskip 1 cm -\frac{1}{2}\delta_{AB}\left(\eta^{ab}\partial_{a}s^{A}\partial_{b}s^{B}\right)-\frac{1}{2}s^{A}s^{B}\partial_{A}\partial_{B}V+\mathcal{O}(\textrm{higher~order~terms})\Bigg]
\nonumber
\\
&=L_{0}+\left[\frac{\kappa}{2}h-\frac{\kappa^{2}}{4}\left(h^{\alpha \beta}h_{\alpha \beta}-\frac{1}{2}h^{2}\right)\right]L_{0}+\frac{\kappa}{2}\delta_{AB}h^{ab}\partial_{a}\phi^{A}\partial_{b}\phi^{B}+\frac{\kappa^{2}}{4}\delta_{AB}hh^{ab}\partial_{a}\phi^{A}\partial_{b}\phi^{B}
\nonumber
\\
&\hskip 1 cm -\frac{\kappa^{2}}{2}\delta_{AB}h^{ac}h^{b}_{c}\partial_{a}\phi^{A}\partial_{b}\phi^{B}-\delta_{AB}\eta^{ab}\partial_{a}\phi^{A}\partial_{b}s^{B}-s^{A}\partial_{A}V+\frac{\kappa}{2}h\left(-\delta_{AB}\eta^{ab}\partial_{a}\phi^{A}\partial_{b}s^{B}-s^{A}\partial_{A}V\right)
\nonumber
\\
&\hskip 1 cm +\delta_{AB}\kappa h^{ab}\partial_{a}\phi^{A}\partial_{b}s^{B}-\frac{1}{2}\delta_{AB}\left(\eta^{ab}\partial_{a}s^{A}\partial_{b}s^{B}\right)-\frac{1}{2}s^{A}s^{B}\partial_{A}\partial_{B}V
\nonumber
\\
&=L_{0}+\left(\frac{\kappa}{2}h L_{0}+\frac{\kappa}{2}\delta_{AB}h^{ab}\partial_{a}\phi^{A}\partial_{b}\phi^{B}-\delta_{AB}\eta^{ab}\partial_{a}\phi^{A}\partial_{b}s^{B}-s^{A}\partial_{A}V\right)+\Bigg[-\frac{\kappa^{2}}{4}\left(h^{\alpha \beta}h_{\alpha \beta}-\frac{1}{2}h^{2}\right)L_{0}
\nonumber
\\
&\hskip 1 cm +\frac{\kappa^{2}}{4}\delta_{AB}\left(hh^{ab}-2h^{ac}h^{b}_{c}\right)\partial_{a}\phi^{A}\partial_{b}\phi^{B}+\frac{\kappa}{2}h\left\{-\delta_{AB}\eta^{ab}\partial_{a}\phi^{A}\partial_{b}s^{B}-s^{A}\partial_{A}V\right\}
\nonumber
\\
&\hskip 1 cm +\delta_{AB}\kappa h^{ab}\partial_{a}\phi^{A}\partial_{b}s^{B}-\frac{1}{2}\delta_{AB}\left\{\eta^{ab}\partial_{a}s^{A}\partial_{b}s^{B}\right\}-\frac{1}{2}s^{A}s^{B}\partial_{A}\partial_{B}V\Bigg]~.
\end{align}
Here we have defined, $L_{0}\equiv-(1/2)\delta_{AB}\left(\eta^{ab}\partial_{a}\phi^{A}\partial_{b}\phi^{B}\right)-V(\phi)$, as the scalar field Lagrangian in flat spacetime. Thus keeping terms quadratic in the gravitational and scalar perturbation, we obtain the following form of the Lagrangian,
\begin{align}
\mathcal{L}^{\rm quadratic}_{\rm matter}&=-\frac{\kappa^{2}}{4}\left(h^{\alpha \beta}h_{\alpha \beta}-\frac{1}{2}h^{2}\right)L_{0}
-\frac{\kappa}{2}h\left(\delta_{AB}\eta^{ab}\partial_{a}\phi^{A}\partial_{b}s^{B}+s^{A}\partial_{A}V\right)
-\frac{1}{2}\delta_{AB}\left(\eta^{ab}\partial_{a}s^{A}\partial_{b}s^{B}\right)
\nonumber
\\
&\hskip 1 cm -\frac{1}{2}s^{A}s^{B}\partial_{A}\partial_{B}V+\frac{\kappa^{2}}{4}\delta_{AB}\left(hh^{ab}-2h^{ac}h^{b}_{c}\right)\partial_{a}\phi^{A}\partial_{b}\phi^{B}+\delta_{AB}\kappa h^{ab}\partial_{a}\phi^{A}\partial_{b}s^{B}~.
\end{align}
Further simplification can be performed by assuming the background scalar field $\phi^{A}$ to be constant. This is consistent with the flat background considered in this work. Therefore the last two terms in the above expression does not contribute and the matter Lagrangian density upto quadratic order in the perturbation can be expressed as,
\begin{align}
\mathcal{L}_{\rm matter}&=-\frac{\kappa^{2}}{4}\left(h^{\alpha \beta}h_{\alpha \beta}-\frac{1}{2}h^{2}\right)L_{0}-\frac{\kappa}{2}h\left(s^{A}\partial_{A}V\right)-\frac{1}{2}\delta_{AB}\left(\eta^{ab}\partial_{a}s^{A}\partial_{b}s^{B}\right)-\frac{1}{2}s^{A}s^{B}\partial_{A}\partial_{B}V
\nonumber
\\
&=h_{ab}P^{ab;cd}\left(\frac{\kappa^{2}}{4}L_{0}\right)h_{cd}-\frac{\kappa}{2}h\left(s^{A}\partial_{A}V\right)+\frac{1}{2}\delta_{AB}s^{A}\square s^{B}-\frac{1}{2}s^{A}s^{B}\partial_{A}\partial_{B}V~.
\end{align}
where in the last line we have neglected some total derivative terms. Note that with the choice $\phi^{A}=\textrm{constant}$, the scalar field Lagrangian for the flat background becomes $-V_{0}$, where $V_{0}$ is the potential associated with the background scalar fields. This is the expression we have used in the main text. 

\section{Expansion of the matter Lagrangian upto quadratic order in the perturbations in scalar-tensor representation}\label{App_B}

Let us consider the matter Lagrangian involving $n$ scalar fields in the scalar-tensor representation of the $f(R)$ gravity and its expansion upto quadratic order in the perturbations, which takes the following form,
\begin{align}
\mathcal{L}_{\rm matter}&=\sqrt{-\bar{g}}\left\{-\frac{1}{2}\Omega^{-2}\delta_{AB}\bar{g}^{ab}\partial_{a}\Phi^{A}\partial_{b}\Phi^{B}-\Omega^{-4}V(\Phi) \right\}
\nonumber
\\
&=\sqrt{-\bar{g}}\left\{-\frac{1}{2}\exp\left(-\frac{\kappa \psi}{\sqrt{6}}\right)\delta_{AB}\bar{g}^{ab}\partial_{a}\Phi^{A}\partial_{b}\Phi^{B}-\exp\left(-2 \frac{\kappa \psi}{\sqrt{6}}\right)V(\Phi) \right\}
\nonumber
\\
&\simeq \sqrt{-\bar{g}}\left[-\frac{1}{2}\delta_{AB}\bar{g}^{ab}\partial_{a}\Phi^{A}\partial_{b}\Phi^{B}-V(\Phi) \right]
+\left(-\frac{\kappa \psi}{\sqrt{6}}\right)\sqrt{-\bar{g}}\left[-\frac{1}{2}\delta_{AB}\bar{g}^{ab}\partial_{a}\Phi^{A}\partial_{b}\Phi^{B}-2V(\Phi) \right]
\nonumber
\\
&\hskip 2 cm +\frac{1}{2}\left(-\frac{\kappa \psi}{\sqrt{6}}\right)^{2}\sqrt{-\bar{g}}\left[-\frac{1}{2}\delta_{AB}\bar{g}^{ab}\partial_{a}\Phi^{A}\partial_{b}\Phi^{B}-4V(\Phi) \right]
\nonumber
\\
&=-\frac{\kappa}{2}hs^{A}\partial_{A}V-\frac{1}{2}\delta_{AB}\eta^{ab}\partial_{a}s^{A}\partial_{b}s^{B}-\frac{1}{2}s^{A}s^{B}\partial_{A}\partial_{B}V
-2\left(-\frac{\kappa \psi}{\sqrt{6}}\right)s^{A}\partial_{A}V
\nonumber
\\
&\hskip 2 cm +\frac{\kappa^{2}}{4}\left(h_{\alpha \beta}h^{\alpha \beta}-\frac{1}{2}h^{2}\right)V_{0}-2\left(-\frac{\kappa \psi}{\sqrt{6}}\right)^{2}V_{0}
-\left(-\frac{\kappa \psi}{\sqrt{6}}\right)\kappa h V_{0}~.
\end{align}
Here we have assumed that the background scalar field $\phi^{A}$ is a constant, such that all the derivatives of $\phi^{A}$ can be set to zero. 
Note that the above quadratic Lagrangian for the matter field depends not only on the scalar perturbation $s^{A}$, but also on the gravitational perturbation $h_{ab}$ and scalar-tensor perturbation $\psi$.  Thus the total Lagrangian involving perturbations upto quadratic order becomes,
\begin{align}
\mathcal{L}_{\rm gr}&+\mathcal{L}_{\rm gf}+\mathcal{L}_{\rm matter}^{\rm scalar-tensor}+\mathcal{L}_{\rm matter}
\nonumber
\\
&=-\frac{1}{2}h_{ab}P^{ab;cd}\square h_{cd}-\frac{\kappa}{2}h\psi \left(\partial W/\partial \psi \right)-\frac{1}{2}\eta^{ab}\partial_{a}\psi \partial_{b}\psi
-\frac{1}{2}\psi^{2}\left(\partial^{2}W/\partial \psi^{2}\right)
\nonumber
\\
&-\frac{\kappa}{2}hs^{A}\partial_{A}V-\frac{1}{2}\delta_{AB}\eta^{ab}\partial_{a}s^{A}\partial_{b}s^{B}-\frac{1}{2}s^{A}s^{B}\partial_{A}\partial_{B}V
-2\left(-\frac{\kappa \psi}{\sqrt{6}}\right)s^{A}\partial_{A}V
\nonumber
\\
&-\frac{\kappa^{2}}{4}\left(h_{\alpha \beta}h^{\alpha \beta}-\frac{1}{2}h^{2}\right)L_{0}+2\left(-\frac{\kappa \psi}{\sqrt{6}}\right)^{2}L_{0}
+\left(-\frac{\kappa \psi}{\sqrt{6}}\right)\kappa h L_{0}~,
\end{align}
which has been used in the main text. 

\section{Scalar potential in the scalar-tensor representation}\label{App_C}

To demonstrate the equivalence of the results derived in the context of scalar-tensor representation with the corresponding results for $f(R)$ gravity we need to evaluate derivatives of the scalar potential $W(\psi)$. This is non-trivial, since the potential is known only an an implicit function of the scalar field $\psi$. In this appendix we will determine the scalar potential and its derivatives with respect to the scalar field, which will be useful in various contexts in this paper. The scalar potential in the conformally transformed frame associated with the scalar field $\psi$ can be read off from \ref{scalar_pot}, which reads,
\begin{align}\label{scal_pot_app}
W(\psi)=\frac{2}{\kappa^{2}}\exp\left(-\frac{2}{\sqrt{6}}\kappa \psi \right)\left\{\chi f'(\chi)-f(\chi)\right\}~;\qquad f'(\chi)=\exp \left(\frac{\kappa \psi}{\sqrt{6}}\right)
\end{align}
Thus taking derivative of the function $W(\psi)$ with respect to $\psi$, we obtain,
\begin{align}
\frac{dW}{d\psi}&=-\frac{4}{\sqrt{6}\kappa}\exp\left(-\frac{2}{\sqrt{6}}\kappa \psi \right)\left\{\chi f'(\chi)-f(\chi)\right\}
+\frac{2}{\kappa^{2}}\exp\left(-\frac{2}{\sqrt{6}}\kappa \psi \right)\chi f''(\chi)\left(\frac{d\chi}{d\psi}\right)~.
\end{align}
Given the relation between $\chi$ and $\psi$ in \ref{scal_pot_app}, we obtain,
\begin{align}
\frac{d\psi}{d\chi}=\frac{\sqrt{6}}{\kappa}\frac{f''(\chi)}{f'(\chi)}~.
\end{align}
Substituting this expression for $(d\psi/d\chi)$ in the expression for $(dW/d\psi)$ derived above we obtain,
\begin{align}
\frac{dW}{d\psi}&=-\frac{4}{\sqrt{6}\kappa}\exp\left(-\frac{2}{\sqrt{6}}\kappa \psi \right)\left\{\chi f'(\chi)-f(\chi)\right\}
+\frac{2}{\kappa^{2}}\exp\left(-\frac{2}{\sqrt{6}}\kappa \psi \right)\chi f''(\chi)\left(\frac{\sqrt{6}}{\kappa}\frac{f''(\chi)}{f'(\chi)}\right)^{-1}
\nonumber
\\
&=-\frac{4}{\sqrt{6}\kappa}\exp\left(-\frac{2}{\sqrt{6}}\kappa \psi \right)\left\{\chi f'(\chi)-f(\chi)\right\}
+\frac{2}{\sqrt{6}\kappa}\exp\left(-\frac{2}{\sqrt{6}}\kappa \psi \right)\chi f'(\chi)~.
\end{align}
Finally, the computation of the second derivative of $W(\psi)$ proceeds along the following lines,
\begin{align}
\frac{d^{2}W}{d\psi^{2}}&=\frac{4}{3}\exp\left(-\frac{2}{\sqrt{6}}\kappa \psi \right)\left\{\chi f'(\chi)-f(\chi)\right\}
-\frac{4}{\sqrt{6}\kappa}\exp\left(-\frac{2}{\sqrt{6}}\kappa \psi \right)\chi f''(\chi)\left(\frac{d\chi}{d\psi}\right)
\nonumber
\\
&-\frac{2}{3}\exp\left(-\frac{2}{\sqrt{6}}\kappa \psi \right)\chi f'(\chi)+\frac{2}{\sqrt{6}\kappa}\exp\left(-\frac{2}{\sqrt{6}}\kappa \psi \right)\left\{f'(\chi)+\chi f''(\chi)\right\}
\left(\frac{d\chi}{d\psi}\right)
\nonumber
\\
&=\exp\left(-\frac{2}{\sqrt{6}}\kappa \psi \right)\left\{\frac{2}{3}\chi f'(\chi)-\frac{4}{3}f(\chi)\right\}
-\frac{2}{3}\exp\left(-\frac{2}{\sqrt{6}}\kappa \psi \right)\chi f'(\chi)
\nonumber
\\
&+\frac{1}{3}\exp\left(-\frac{2}{\sqrt{6}}\kappa \psi \right)\left\{\frac{f'(\chi)^{2}}{f''(\chi)}+\chi f'(\chi)\right\}
\nonumber
\\
&=-\frac{4}{3}f(\chi)\exp\left(-\frac{2}{\sqrt{6}}\kappa \psi \right)+\frac{1}{3}\exp\left(-\frac{2}{\sqrt{6}}\kappa \psi \right)\left\{\frac{f'(\chi)^{2}}{f''(\chi)}+\chi f'(\chi)\right\}~.
\end{align}
This expression when evaluated for the background spacetime, where $\psi=0$ and on-shell $\chi=R=0$. Thus we obtain the above second derivative term to yield,,
\begin{align}
\frac{d^{2}W}{d\psi^{2}}&=-\frac{4}{3}f(R=0)+\frac{1}{3}\frac{f'(R=0)^{2}}{f''(R=0)}
\end{align}
From our consideration of $f(R)$ gravity, it follows that $f(R=0)=0$ and $f'(R=0)=1$, which yields, $(d^{2}W/d\psi^{2})\sim f''(R=0)^{-1}$. This result has been used in the main text.  
\bibliography{References}

\providecommand{\href}[2]{#2}\begingroup\raggedright\begin{thebibliography}{10}

\bibitem{Aad:2012tfa}
{\bfseries ATLAS} Collaboration, G.~Aad {\em et~al.}, ``{Observation of a new
  particle in the search for the Standard Model Higgs boson with the ATLAS
  detector at the LHC},''
  \href{http://dx.doi.org/10.1016/j.physletb.2012.08.020}{{\em Phys. Lett. B}
  {\bfseries 716} (2012) 1--29},
  \href{http://arxiv.org/abs/1207.7214}{{\ttfamily arXiv:1207.7214 [hep-ex]}}.

\bibitem{Chatrchyan:2012ufa}
{\bfseries CMS} Collaboration, S.~Chatrchyan {\em et~al.}, ``{Observation of a
  New Boson at a Mass of 125 GeV with the CMS Experiment at the LHC},''
  \href{http://dx.doi.org/10.1016/j.physletb.2012.08.021}{{\em Phys. Lett. B}
  {\bfseries 716} (2012) 30--61},
  \href{http://arxiv.org/abs/1207.7235}{{\ttfamily arXiv:1207.7235 [hep-ex]}}.

\bibitem{Spira:1995rr}
M.~Spira, A.~Djouadi, D.~Graudenz, and P.~Zerwas, ``{Higgs boson production at
  the LHC},'' \href{http://dx.doi.org/10.1016/0550-3213(95)00379-7}{{\em Nucl.
  Phys. B} {\bfseries 453} (1995) 17--82},
  \href{http://arxiv.org/abs/hep-ph/9504378}{{\ttfamily arXiv:hep-ph/9504378}}.

\bibitem{DeRujula:2010ys}
A.~De~Rujula, J.~Lykken, M.~Pierini, C.~Rogan, and M.~Spiropulu, ``{Higgs
  Look-Alikes at the LHC},''
  \href{http://dx.doi.org/10.1103/PhysRevD.82.013003}{{\em Phys. Rev. D}
  {\bfseries 82} (2010) 013003},
  \href{http://arxiv.org/abs/1001.5300}{{\ttfamily arXiv:1001.5300 [hep-ph]}}.

\bibitem{Weinberg:1967tq}
S.~Weinberg, ``{A Model of Leptons},''
  \href{http://dx.doi.org/10.1103/PhysRevLett.19.1264}{{\em Phys. Rev. Lett.}
  {\bfseries 19} (1967) 1264--1266}.

\bibitem{Salam:1968rm}
A.~Salam, ``{Weak and Electromagnetic Interactions},''
  \href{http://dx.doi.org/10.1142/9789812795915\_0034}{{\em Conf. Proc. C}
  {\bfseries 680519} (1968) 367--377}.

\bibitem{Cornwall:1973tb}
J.~M. Cornwall, D.~N. Levin, and G.~Tiktopoulos, ``{Uniqueness of spontaneously
  broken gauge theories},''
  \href{http://dx.doi.org/10.1103/PhysRevLett.30.1268}{{\em Phys. Rev. Lett.}
  {\bfseries 30} (1973) 1268--1270}. [Erratum: Phys.Rev.Lett. 31, 572 (1973)].

\bibitem{Barate:2003sz}
{\bfseries LEP Working Group for Higgs boson searches, ALEPH, DELPHI, L3, OPAL}
  Collaboration, R.~Barate {\em et~al.}, ``{Search for the standard model Higgs
  boson at LEP},'' \href{http://dx.doi.org/10.1016/S0370-2693(03)00614-2}{{\em
  Phys. Lett. B} {\bfseries 565} (2003) 61--75},
  \href{http://arxiv.org/abs/hep-ex/0306033}{{\ttfamily arXiv:hep-ex/0306033}}.

\bibitem{Evans:2008zzb}
``{LHC Machine},'' \href{http://dx.doi.org/10.1088/1748-0221/3/08/S08001}{{\em
  JINST} {\bfseries 3} (2008) S08001}.

\bibitem{Susskind:1982mw}
L.~Susskind, ``{THE GAUGE HIERARCHY PROBLEM, TECHNICOLOR, SUPERSYMMETRY, AND
  ALL THAT.},'' \href{http://dx.doi.org/10.1016/0370-1573(84)90208-4}{{\em
  Phys. Rept.} {\bfseries 104} (1984) 181--193}.

\bibitem{Martin:1997ns}
S.~P. Martin, {\em {A Supersymmetry primer}}, vol.~21,
  \href{http://dx.doi.org/10.1142/9789812839657\_0001}{pp.~1--153}.
\newblock 2010.
\newblock \href{http://arxiv.org/abs/hep-ph/9709356}{{\ttfamily
  arXiv:hep-ph/9709356}}.

\bibitem{Freedman:2012zz}
D.~Z. Freedman and A.~Van~Proeyen, {\em {Supergravity}}.
\newblock Cambridge Univ. Press, Cambridge, UK, 5, 2012.

\bibitem{Appelquist:1997fp}
T.~Appelquist, J.~Terning, and L.~Wijewardhana, ``{Postmodern technicolor},''
  \href{http://dx.doi.org/10.1103/PhysRevLett.79.2767}{{\em Phys. Rev. Lett.}
  {\bfseries 79} (1997) 2767--2770},
  \href{http://arxiv.org/abs/hep-ph/9706238}{{\ttfamily arXiv:hep-ph/9706238}}.

\bibitem{Hill:2002ap}
C.~T. Hill and E.~H. Simmons, ``{Strong Dynamics and Electroweak Symmetry
  Breaking},'' \href{http://dx.doi.org/10.1016/S0370-1573(03)00140-6}{{\em
  Phys. Rept.} {\bfseries 381} (2003) 235--402},
  \href{http://arxiv.org/abs/hep-ph/0203079}{{\ttfamily arXiv:hep-ph/0203079}}.
  [Erratum: Phys.Rept. 390, 553--554 (2004)].

\bibitem{Randall:1999ee}
L.~Randall and R.~Sundrum, ``{A Large mass hierarchy from a small extra
  dimension},'' \href{http://dx.doi.org/10.1103/PhysRevLett.83.3370}{{\em Phys.
  Rev. Lett.} {\bfseries 83} (1999) 3370--3373},
  \href{http://arxiv.org/abs/hep-ph/9905221}{{\ttfamily arXiv:hep-ph/9905221}}.

\bibitem{ArkaniHamed:1998rs}
N.~Arkani-Hamed, S.~Dimopoulos, and G.~Dvali, ``{The Hierarchy problem and new
  dimensions at a millimeter},''
  \href{http://dx.doi.org/10.1016/S0370-2693(98)00466-3}{{\em Phys. Lett. B}
  {\bfseries 429} (1998) 263--272},
  \href{http://arxiv.org/abs/hep-ph/9803315}{{\ttfamily arXiv:hep-ph/9803315}}.

\bibitem{Antoniadis:1990ew}
I.~Antoniadis, ``{A Possible new dimension at a few TeV},''
  \href{http://dx.doi.org/10.1016/0370-2693(90)90617-F}{{\em Phys. Lett. B}
  {\bfseries 246} (1990) 377--384}.

\bibitem{Autermann:2017chm}
{\bfseries ATLAS, CMS} Collaboration, C.~Autermann, ``{Search for supersymmetry
  at the LHC},'' \href{http://dx.doi.org/10.1051/epjconf/201716401028}{{\em EPJ
  Web Conf.} {\bfseries 164} (2017) 01028}.

\bibitem{Alwall:2008ag}
J.~Alwall, P.~Schuster, and N.~Toro, ``{Simplified Models for a First
  Characterization of New Physics at the LHC},''
  \href{http://dx.doi.org/10.1103/PhysRevD.79.075020}{{\em Phys. Rev. D}
  {\bfseries 79} (2009) 075020},
  \href{http://arxiv.org/abs/0810.3921}{{\ttfamily arXiv:0810.3921 [hep-ph]}}.

\bibitem{Khachatryan:2016epu}
{\bfseries CMS} Collaboration, V.~Khachatryan {\em et~al.}, ``{Inclusive search
  for supersymmetry using razor variables in pp collisions at $\sqrt s=$ 13
  TeV},'' \href{http://dx.doi.org/10.1103/PhysRevD.95.012003}{{\em Phys. Rev.
  D} {\bfseries 95} no.~1, (2017) 012003},
  \href{http://arxiv.org/abs/1609.07658}{{\ttfamily arXiv:1609.07658
  [hep-ex]}}.

\bibitem{Autermann:2016les}
C.~Autermann, ``{Experimental status of supersymmetry after the LHC Run-I},''
  \href{http://dx.doi.org/10.1016/j.ppnp.2016.06.001}{{\em Prog. Part. Nucl.
  Phys.} {\bfseries 90} (2016) 125--155},
  \href{http://arxiv.org/abs/1609.01686}{{\ttfamily arXiv:1609.01686
  [hep-ex]}}.

\bibitem{Shmatov:2007mg}
{\bfseries ATLAS, CMS} Collaboration, S.~Shmatov, ``{Search for Extra
  Dimensions with Atlas and CMS Detectors at the LHC},''
  \href{http://dx.doi.org/10.1142/9789812790873\_0246}{{\em Conf. Proc. C}
  {\bfseries 060726} (2006) 1141--1145},
  \href{http://arxiv.org/abs/0707.0470}{{\ttfamily arXiv:0707.0470 [hep-ex]}}.

\bibitem{Chatrchyan:2011fq}
{\bfseries CMS} Collaboration, S.~Chatrchyan {\em et~al.}, ``{Search for
  signatures of extra dimensions in the diphoton mass spectrum at the Large
  Hadron Collider},''
  \href{http://dx.doi.org/10.1103/PhysRevLett.108.111801}{{\em Phys. Rev.
  Lett.} {\bfseries 108} (2012) 111801},
  \href{http://arxiv.org/abs/1112.0688}{{\ttfamily arXiv:1112.0688 [hep-ex]}}.

\bibitem{Belyaev:2013ida}
A.~Belyaev, M.~S. Brown, R.~Foadi, and M.~T. Frandsen, ``{The Technicolor Higgs
  in the Light of LHC Data},''
  \href{http://dx.doi.org/10.1103/PhysRevD.90.035012}{{\em Phys. Rev. D}
  {\bfseries 90} (2014) 035012},
  \href{http://arxiv.org/abs/1309.2097}{{\ttfamily arXiv:1309.2097 [hep-ph]}}.

\bibitem{Meissner:2007xv}
K.~A. Meissner and H.~Nicolai, ``{Effective action, conformal anomaly and the
  issue of quadratic divergences},''
  \href{http://dx.doi.org/10.1016/j.physletb.2007.12.035}{{\em Phys. Lett. B}
  {\bfseries 660} (2008) 260--266},
  \href{http://arxiv.org/abs/0710.2840}{{\ttfamily arXiv:0710.2840 [hep-th]}}.

\bibitem{Chankowski:2014fva}
P.~H. Chankowski, A.~Lewandowski, K.~A. Meissner, and H.~Nicolai, ``{Softly
  broken conformal symmetry and the stability of the electroweak scale},''
  \href{http://dx.doi.org/10.1142/S0217732315500066}{{\em Mod. Phys. Lett. A}
  {\bfseries 30} no.~02, (2015) 1550006},
  \href{http://arxiv.org/abs/1404.0548}{{\ttfamily arXiv:1404.0548 [hep-ph]}}.

\bibitem{Meissner:2018mvq}
K.~A. Meissner, H.~Nicolai, and J.~Plefka, ``{Softly broken conformal symmetry
  with quantum gravitational corrections},''
  \href{http://dx.doi.org/10.1016/j.physletb.2019.01.066}{{\em Phys. Lett. B}
  {\bfseries 791} (2019) 62--65},
  \href{http://arxiv.org/abs/1811.05216}{{\ttfamily arXiv:1811.05216
  [hep-th]}}.

\bibitem{Loebbert:2018xsd}
F.~Loebbert, J.~Miczajka, and J.~Plefka, ``{Consistent Conformal Extensions of
  the Standard Model},''
  \href{http://dx.doi.org/10.1103/PhysRevD.99.015026}{{\em Phys. Rev. D}
  {\bfseries 99} no.~1, (2019) 015026},
  \href{http://arxiv.org/abs/1805.09727}{{\ttfamily arXiv:1805.09727
  [hep-ph]}}.

\bibitem{Kwapisz:2017vjt}
J.~Kwapisz and K.~A. Meissner, ``{Conformal Standard Model and Inflation},''
  \href{http://dx.doi.org/10.5506/APhysPolB.49.115}{{\em Acta Phys. Polon. B}
  {\bfseries 49} (2018) 115}, \href{http://arxiv.org/abs/1712.03778}{{\ttfamily
  arXiv:1712.03778 [gr-qc]}}.

\bibitem{Lewandowski:2017wov}
A.~Lewandowski, K.~A. Meissner, and H.~Nicolai, ``{Conformal Standard Model,
  Leptogenesis and Dark Matter},''
  \href{http://dx.doi.org/10.1103/PhysRevD.97.035024}{{\em Phys. Rev. D}
  {\bfseries 97} no.~3, (2018) 035024},
  \href{http://arxiv.org/abs/1710.06149}{{\ttfamily arXiv:1710.06149
  [hep-ph]}}.

\bibitem{Latosinski:2015pba}
A.~Latosinski, A.~Lewandowski, K.~A. Meissner, and H.~Nicolai, ``{Conformal
  Standard Model with an extended scalar sector},''
  \href{http://dx.doi.org/10.1007/JHEP10(2015)170}{{\em JHEP} {\bfseries 10}
  (2015) 170}, \href{http://arxiv.org/abs/1507.01755}{{\ttfamily
  arXiv:1507.01755 [hep-ph]}}.

\bibitem{Hamada:2012bp}
Y.~Hamada, H.~Kawai, and K.-y. Oda, ``{Bare Higgs mass at Planck scale},''
  \href{http://dx.doi.org/10.1103/PhysRevD.87.053009}{{\em Phys. Rev. D}
  {\bfseries 87} no.~5, (2013) 053009},
  \href{http://arxiv.org/abs/1210.2538}{{\ttfamily arXiv:1210.2538 [hep-ph]}}.
  [Erratum: Phys.Rev.D 89, 059901 (2014)].

\bibitem{Meissner:2009gs}
K.~A. Meissner and H.~Nicolai, ``{Conformal invariance from non-conformal
  gravity},'' \href{http://dx.doi.org/10.1103/PhysRevD.80.086005}{{\em Phys.
  Rev. D} {\bfseries 80} (2009) 086005},
  \href{http://arxiv.org/abs/0907.3298}{{\ttfamily arXiv:0907.3298 [hep-th]}}.

\bibitem{Antipin:2013exa}
O.~Antipin, M.~Mojaza, and F.~Sannino, ``{Conformal Extensions of the Standard
  Model with Veltman Conditions},''
  \href{http://dx.doi.org/10.1103/PhysRevD.89.085015}{{\em Phys. Rev. D}
  {\bfseries 89} no.~8, (2014) 085015},
  \href{http://arxiv.org/abs/1310.0957}{{\ttfamily arXiv:1310.0957 [hep-ph]}}.

\bibitem{Woodard:2015zca}
R.~P. Woodard, ``{Ostrogradsky's theorem on Hamiltonian instability},''
  \href{http://dx.doi.org/10.4249/scholarpedia.32243}{{\em Scholarpedia}
  {\bfseries 10} no.~8, (2015) 32243},
  \href{http://arxiv.org/abs/1506.02210}{{\ttfamily arXiv:1506.02210
  [hep-th]}}.

\bibitem{Sotiriou:2008rp}
T.~P. Sotiriou and V.~Faraoni, ``{f(R) Theories Of Gravity},''
  \href{http://dx.doi.org/10.1103/RevModPhys.82.451}{{\em Rev. Mod. Phys.}
  {\bfseries 82} (2010) 451--497},
  \href{http://arxiv.org/abs/0805.1726}{{\ttfamily arXiv:0805.1726 [gr-qc]}}.

\bibitem{Nojiri:2010wj}
S.~Nojiri and S.~D. Odintsov, ``{Unified cosmic history in modified gravity:
  from F(R) theory to Lorentz non-invariant models},''
  \href{http://dx.doi.org/10.1016/j.physrep.2011.04.001}{{\em Phys. Rept.}
  {\bfseries 505} (2011) 59--144},
  \href{http://arxiv.org/abs/1011.0544}{{\ttfamily arXiv:1011.0544 [gr-qc]}}.

\bibitem{DeFelice:2010aj}
A.~De~Felice and S.~Tsujikawa, ``{f(R) theories},''
  \href{http://dx.doi.org/10.12942/lrr-2010-3}{{\em Living Rev. Rel.}
  {\bfseries 13} (2010) 3}, \href{http://arxiv.org/abs/1002.4928}{{\ttfamily
  arXiv:1002.4928 [gr-qc]}}.

\bibitem{Chakraborty:2020ktp}
S.~Chakraborty and S.~Ghosh, ``{Non-trivial ground state for gravitational
  perturbation in quadratic gravity},''
  \href{http://arxiv.org/abs/2001.04680}{{\ttfamily arXiv:2001.04680 [gr-qc]}}.

\bibitem{Chakraborty:2016gpg}
S.~Chakraborty and S.~SenGupta, ``{Gravity stabilizes itself},''
  \href{http://dx.doi.org/10.1140/epjc/s10052-017-5138-5}{{\em Eur. Phys. J. C}
  {\bfseries 77} no.~8, (2017) 573},
  \href{http://arxiv.org/abs/1701.01032}{{\ttfamily arXiv:1701.01032 [gr-qc]}}.

\bibitem{Carames:2012gr}
T.~Carames, M.~Guimaraes, and J.~Hoff~da Silva, ``{Effective gravitational
  equations for $f(R)$ braneworld models},''
  \href{http://dx.doi.org/10.1103/PhysRevD.87.106011}{{\em Phys. Rev. D}
  {\bfseries 87} no.~10, (2013) 106011},
  \href{http://arxiv.org/abs/1205.4980}{{\ttfamily arXiv:1205.4980 [gr-qc]}}.

\bibitem{Haghani:2012zq}
Z.~Haghani, H.~R. Sepangi, and S.~Shahidi, ``{Cosmological dynamics of brane
  f(R) gravity},'' \href{http://dx.doi.org/10.1088/1475-7516/2012/02/031}{{\em
  JCAP} {\bfseries 02} (2012) 031},
  \href{http://arxiv.org/abs/1201.6448}{{\ttfamily arXiv:1201.6448 [gr-qc]}}.

\bibitem{Chakraborty:2015taq}
S.~Chakraborty and S.~SenGupta, ``{Spherically symmetric brane in a bulk of
  $f(R)$ and Gauss--Bonnet gravity},''
  \href{http://dx.doi.org/10.1088/0264-9381/33/22/225001}{{\em Class. Quant.
  Grav.} {\bfseries 33} no.~22, (2016) 225001},
  \href{http://arxiv.org/abs/1510.01953}{{\ttfamily arXiv:1510.01953 [gr-qc]}}.

\bibitem{Chakraborty:2014xla}
S.~Chakraborty and S.~SenGupta, ``{Spherically symmetric brane spacetime with
  bulk $f(\mathcal {R})$ gravity},''
  \href{http://dx.doi.org/10.1140/epjc/s10052-014-3234-3}{{\em Eur. Phys. J. C}
  {\bfseries 75} no.~1, (2015) 11},
  \href{http://arxiv.org/abs/1409.4115}{{\ttfamily arXiv:1409.4115 [gr-qc]}}.

\bibitem{Padmanabhan:2013xyr}
T.~Padmanabhan and D.~Kothawala, ``{Lanczos-Lovelock models of gravity},''
  \href{http://dx.doi.org/10.1016/j.physrep.2013.05.007}{{\em Phys. Rept.}
  {\bfseries 531} (2013) 115--171},
  \href{http://arxiv.org/abs/1302.2151}{{\ttfamily arXiv:1302.2151 [gr-qc]}}.

\bibitem{Dadhich:2008df}
N.~Dadhich, ``{Characterization of the Lovelock gravity by Bianchi
  derivative},'' \href{http://dx.doi.org/10.1007/s12043-010-0080-1}{{\em
  Pramana} {\bfseries 74} (2010) 875--882},
  \href{http://arxiv.org/abs/0802.3034}{{\ttfamily arXiv:0802.3034 [gr-qc]}}.

\bibitem{Kastor:2012se}
D.~Kastor, ``{The Riemann-Lovelock Curvature Tensor},''
  \href{http://dx.doi.org/10.1088/0264-9381/29/15/155007}{{\em Class. Quant.
  Grav.} {\bfseries 29} (2012) 155007},
  \href{http://arxiv.org/abs/1202.5287}{{\ttfamily arXiv:1202.5287 [hep-th]}}.

\bibitem{Chakraborty:2015wma}
S.~Chakraborty, ``{Lanczos-Lovelock gravity from a thermodynamic
  perspective},'' \href{http://dx.doi.org/10.1007/JHEP08(2015)029}{{\em JHEP}
  {\bfseries 08} (2015) 029}, \href{http://arxiv.org/abs/1505.07272}{{\ttfamily
  arXiv:1505.07272 [gr-qc]}}.

\bibitem{Charmousis:2014mia}
C.~Charmousis, ``{From Lovelock to Horndeski`s Generalized Scalar Tensor
  Theory},'' \href{http://dx.doi.org/10.1007/978-3-319-10070-8\_2}{{\em Lect.
  Notes Phys.} {\bfseries 892} (2015) 25--56},
  \href{http://arxiv.org/abs/1405.1612}{{\ttfamily arXiv:1405.1612 [gr-qc]}}.

\bibitem{Barausse:2015wia}
E.~Barausse and K.~Yagi, ``{Gravitation-Wave Emission in Shift-Symmetric
  Horndeski Theories},''
  \href{http://dx.doi.org/10.1103/PhysRevLett.115.211105}{{\em Phys. Rev.
  Lett.} {\bfseries 115} no.~21, (2015) 211105},
  \href{http://arxiv.org/abs/1509.04539}{{\ttfamily arXiv:1509.04539 [gr-qc]}}.

\bibitem{Babichev:2016rlq}
E.~Babichev, C.~Charmousis, and A.~Lehébel, ``{Black holes and stars in
  Horndeski theory},''
  \href{http://dx.doi.org/10.1088/0264-9381/33/15/154002}{{\em Class. Quant.
  Grav.} {\bfseries 33} no.~15, (2016) 154002},
  \href{http://arxiv.org/abs/1604.06402}{{\ttfamily arXiv:1604.06402 [gr-qc]}}.

\bibitem{Bhattacharya:2016naa}
S.~Bhattacharya and S.~Chakraborty, ``{Constraining some Horndeski gravity
  theories},'' \href{http://dx.doi.org/10.1103/PhysRevD.95.044037}{{\em Phys.
  Rev. D} {\bfseries 95} no.~4, (2017) 044037},
  \href{http://arxiv.org/abs/1607.03693}{{\ttfamily arXiv:1607.03693 [gr-qc]}}.

\bibitem{Mukherjee:2017fqz}
S.~Mukherjee and S.~Chakraborty, ``{Horndeski theories confront the Gravity
  Probe B experiment},''
  \href{http://dx.doi.org/10.1103/PhysRevD.97.124007}{{\em Phys. Rev. D}
  {\bfseries 97} no.~12, (2018) 124007},
  \href{http://arxiv.org/abs/1712.00562}{{\ttfamily arXiv:1712.00562 [gr-qc]}}.

\bibitem{Cognola:2005de}
G.~Cognola, E.~Elizalde, S.~Nojiri, S.~D. Odintsov, and S.~Zerbini, ``{One-loop
  f(R) gravity in de Sitter universe},''
  \href{http://dx.doi.org/10.1088/1475-7516/2005/02/010}{{\em JCAP} {\bfseries
  02} (2005) 010}, \href{http://arxiv.org/abs/hep-th/0501096}{{\ttfamily
  arXiv:hep-th/0501096}}.

\bibitem{Bamba:2014mua}
K.~Bamba, G.~Cognola, S.~D. Odintsov, and S.~Zerbini, ``{One-loop modified
  gravity in a de Sitter universe, quantum-corrected inflation, and its
  confrontation with the Planck result},''
  \href{http://dx.doi.org/10.1103/PhysRevD.90.023525}{{\em Phys. Rev. D}
  {\bfseries 90} no.~2, (2014) 023525},
  \href{http://arxiv.org/abs/1404.4311}{{\ttfamily arXiv:1404.4311 [gr-qc]}}.

\bibitem{Bamba:2015uxa}
K.~Bamba, S.~D. Odintsov, and P.~V. Tretyakov, ``{Inflation in a
  conformally-invariant two-scalar-field theory with an extra $R^2$ term},''
  \href{http://dx.doi.org/10.1140/epjc/s10052-015-3565-8}{{\em Eur. Phys. J. C}
  {\bfseries 75} no.~7, (2015) 344},
  \href{http://arxiv.org/abs/1505.00854}{{\ttfamily arXiv:1505.00854
  [hep-th]}}.

\bibitem{Elizalde:2017mrn}
E.~Elizalde, S.~Odintsov, L.~Sebastiani, and R.~Myrzakulov, ``{Beyond-one-loop
  quantum gravity action yielding both inflation and late-time acceleration},''
  \href{http://dx.doi.org/10.1016/j.nuclphysb.2017.06.003}{{\em Nucl. Phys. B}
  {\bfseries 921} (2017) 411--435},
  \href{http://arxiv.org/abs/1706.01879}{{\ttfamily arXiv:1706.01879 [gr-qc]}}.

\bibitem{Buchbinder:1992rb}
I.~L. Buchbinder, S.~D. Odintsov, and I.~L. Shapiro, {\em {Effective action in
  quantum gravity}}.
\newblock
1992.
\newblock

\bibitem{Sotiriou:2006hs}
T.~P. Sotiriou, ``{f(R) gravity and scalar-tensor theory},''
  \href{http://dx.doi.org/10.1088/0264-9381/23/17/003}{{\em Class. Quant.
  Grav.} {\bfseries 23} (2006) 5117--5128},
  \href{http://arxiv.org/abs/gr-qc/0604028}{{\ttfamily arXiv:gr-qc/0604028}}.

\bibitem{Briscese:2006xu}
F.~Briscese, E.~Elizalde, S.~Nojiri, and S.~Odintsov, ``{Phantom scalar dark
  energy as modified gravity: Understanding the origin of the Big Rip
  singularity},'' \href{http://dx.doi.org/10.1016/j.physletb.2007.01.013}{{\em
  Phys. Lett. B} {\bfseries 646} (2007) 105--111},
  \href{http://arxiv.org/abs/hep-th/0612220}{{\ttfamily arXiv:hep-th/0612220}}.

\bibitem{Chakraborty:2016ydo}
S.~Chakraborty and S.~SenGupta, ``{Solving higher curvature gravity
  theories},'' \href{http://dx.doi.org/10.1140/epjc/s10052-016-4394-0}{{\em
  Eur. Phys. J. C} {\bfseries 76} no.~10, (2016) 552},
  \href{http://arxiv.org/abs/1604.05301}{{\ttfamily arXiv:1604.05301 [gr-qc]}}.

\end{thebibliography}\endgroup

\bibliographystyle{./utphys1}
\end{document}